\documentclass[twocolumn,amsfonts,showpacs,superscriptaddress,nofootinbib]{revtex4-1}
\usepackage[dvipdfmx]{graphicx}
\usepackage{amssymb,amsmath,amsthm,booktabs,mathtools}
\usepackage{bm}
\usepackage{color}

\usepackage{tikz}
\usepackage{pgfplots}
\usepackage{enumitem}
\usepackage{units}

\usepackage{amsmath}

\newcommand{\beq}{\begin{equation}}
\newcommand{\eeq}{\end{equation}}
\def\frc#1#2{\frac{#1}{#2}}
\def\lt#1{\left#1}
\def\rt#1{\right#1}
\def\t#1{\tilde{#1}}
\newcommand{\p}{\partial}
\newcommand{\ep}{\epsilon}

\begin{document}

\title{Generalized Hydrodynamics on an Atom Chip}

\author{M. Schemmer}
\affiliation
{Laboratoire  Charles  Fabry,  Institut  d'Optique,  CNRS,  Universit\'e Paris  Saclay, 2  Avenue  Augustin  Fresnel,  F-91127  Palaiseau  Cedex,  France}

\author{I. Bouchoule}
\affiliation
{Laboratoire  Charles  Fabry,  Institut  d'Optique,  CNRS,  Universit\'e Paris  Saclay, 2  Avenue  Augustin  Fresnel,  F-91127  Palaiseau  Cedex,  France}

\author{B. Doyon}
\affiliation
{Department of Mathematics, King's College London, Strand, London WC2R 2LS, UK
}
\author{J. Dubail}
\affiliation
{Laboratoire de Physique et Chimie Th\'eoriques, CNRS, Universit\'e de Lorraine, UMR 7019, F-54506 Vandoeuvre-les-Nancy, France
}

\begin{abstract}
The emergence of a special type of fluid-like behavior at large scales in one-dimensional (1d) quantum integrable systems, theoretically predicted in 2016, is established experimentally, by monitoring the time evolution of the insitu density profile of a single 1d cloud of $^{87}{\rm Rb}$ atoms trapped on
an atom chip after a quench of the longitudinal trapping potential. The theory can be viewed as a dynamical extension of the thermodynamics of Yang and Yang, and applies to the whole range of repulsive interaction strength and temperature of the gas. The measurements, performed on weakly interacting atomic clouds that lie at the crossover between the quasicondensate and the ideal Bose gas regimes, are in very good agreement with the 2016 theory. This contrasts with the previously existing ``conventional'' hydrodynamic approach---that relies on the assumption of local thermal equilibrium---, which is unable to reproduce the experimental data.
 \end{abstract}

\maketitle

The emergent hydrodynamic behavior of many interacting particles is a fascinating phenomenon: at the atomic level, all classical/quantum systems are described by the Newton/Schr\"odinger equation, yet these unique microscopic descriptions give rise to a wealth of different liquid and gas phases at larger scales, from the ideal gas to liquid water to plasmas to superfluid helium to Bose-Einstein condensates, to name but a few. Inferring the correct hydrodynamic behavior directly from the microscopic constituents of a many-body system is, in general, a very ambitious task that typically involves extensive numerical simulations and a hierarchy of different modelings on intermediate scales \cite{hanssen,allen}.

However, there exist a few simpler systems where the emergence of a special kind of hydrodynamics can be linked directly to the underlying microscopic rules \cite{spohn_book}. One such system is the {\it one-dimensional (1d) classical billiard}, or {\it hard-rod gas} \cite{spohn_book,percus, dobrods}, whose hydrodynamic behavior, as seen below, is similar 
to that of the quantum system studied in this paper. The hard-rod gas consists of $N$ identical impenetrable rods of fixed diameter $\Delta$ that move along a line, and exchange their momenta upon colliding elastically. At large $N$, the 1d billiard admits a hydrodynamic description: in the limit of density variations of very long wavelength, the billiard can be described by a continuous distribution $\rho(x, v)$ of rods moving at velocity $v$ around a position $x$ and this distribution satisfies an exact evolution equation which resembles the Liouville equation for phase space densities, up to a renormalization of the bare velocity $v$ (see Eq. (\ref{eq:ghd}) below). 
The latter renormalization encodes the following microscopic mechanism: when one rod with velocity $v$ hits another one with velocity $w < v$ from the left, they exchange their momenta. Equivalently, because all the rods are identical, one can think of the collision as an instantaneous exchange of their positions, as if the rod with bare velocity $v$ jumped instantaneously by a distance $\Delta$ to the right. Thus, the time needed by that rod to travel a distance $\ell$ is not $\ell/v$, but rather $(\ell - \Delta)/v$. 
For a finite density of rods, this results in each rod with bare velocity $v$ moving at an {\it effective velocity} $v^{\rm eff} (v)$, that depends on the distribution $\rho(x,v)$ \cite{spohn_book,percus,dobrods}. In distributions of long wavelengths,  the evolution equation for $\rho(x,v)$ becomes a hydrodynamic flow controlled by the local effective velocity $v^{\rm eff}$. The 1d billiard thus exhibits an interesting hydrodynamic behavior that is straightforwardly related to its microscopics.
Slight generalizations of that model exist, where the jumping distance $\Delta$ depends on the relative velocity $v-w$, that possess a similar hydrodynamic description \cite{dyc}.

\begin{figure}[t]
	\begin{tikzpicture}
		\draw (0,0) node{\includegraphics[width=0.3\textwidth]{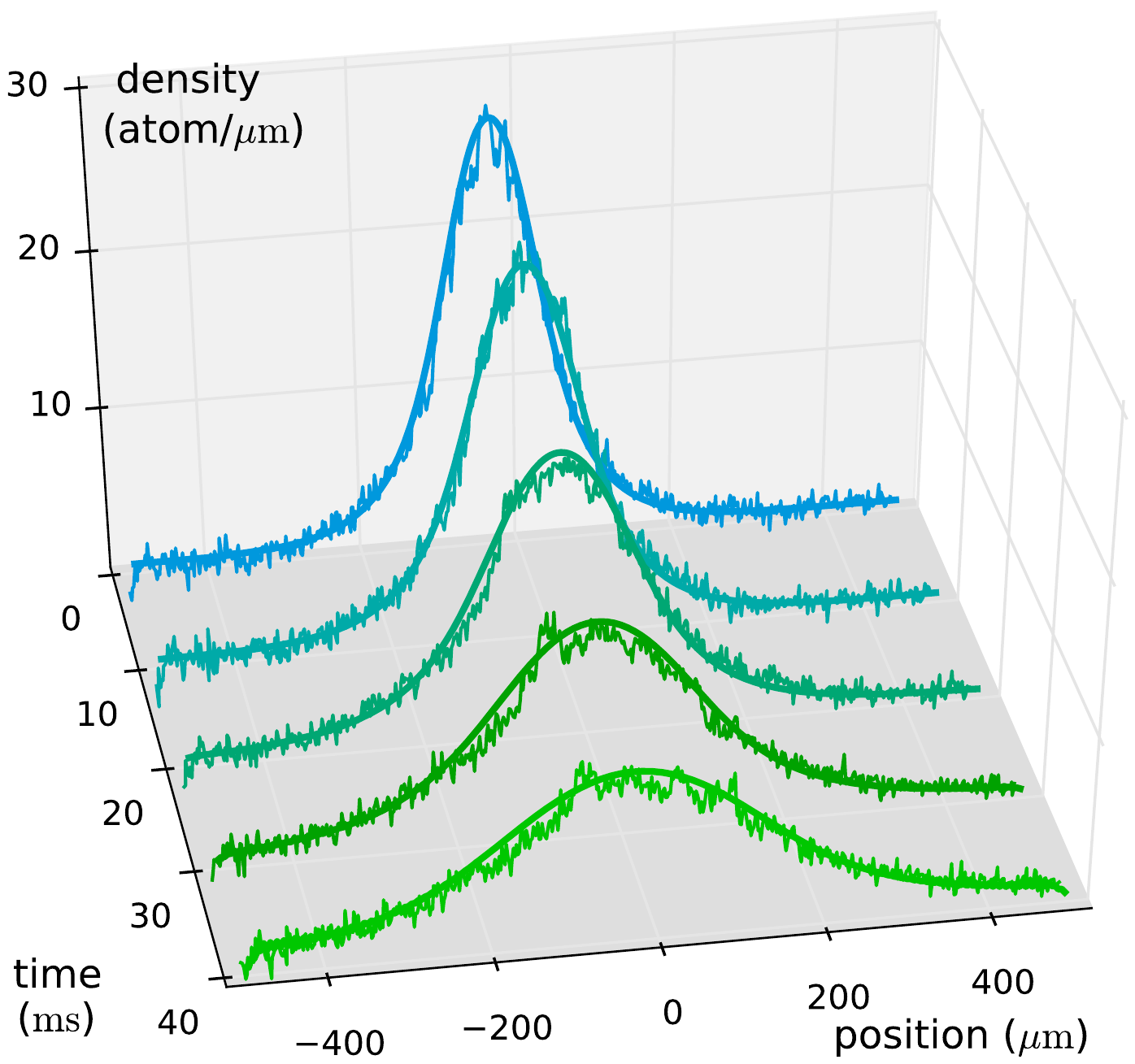}};
		\draw (4.4,-1.3) node{\includegraphics[width=0.16\textwidth]{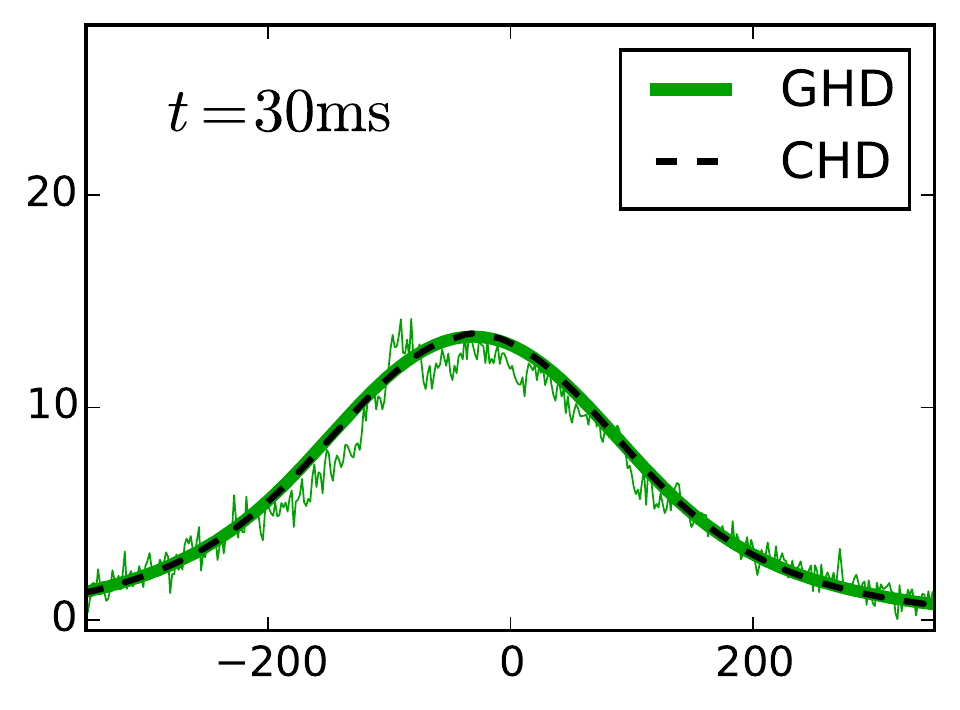}};
		\draw (4.4,1.3) node{\includegraphics[width=0.16\textwidth]{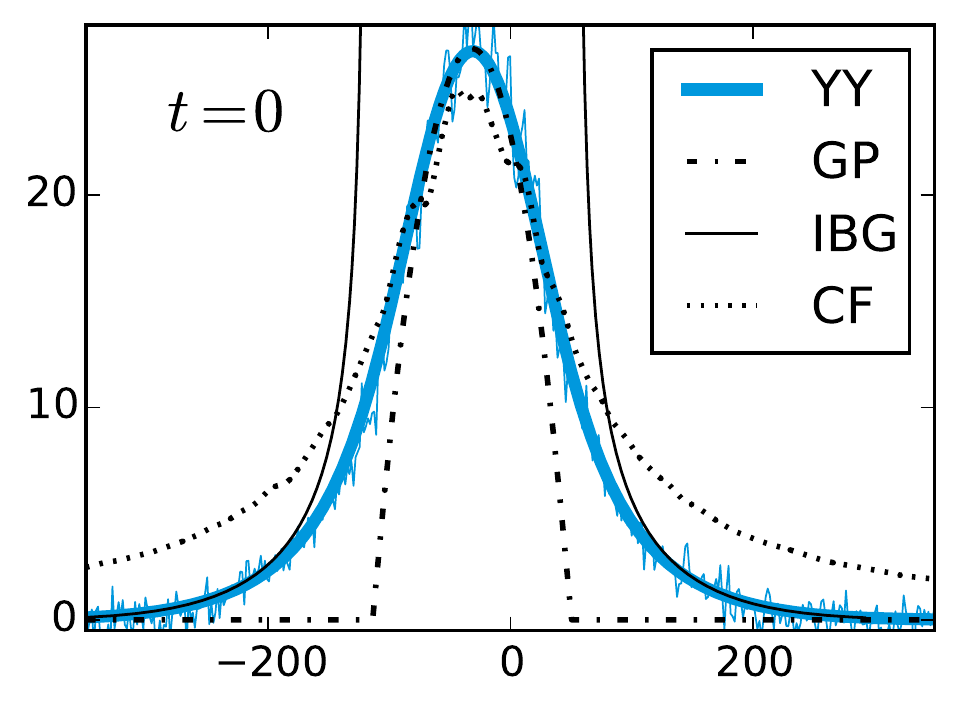}};
		\draw (-2.6,-2.6) node{{\small $(i)$}};
		\draw (3.2,0.1) node{{\small $(ii)$}};
		\draw (3.2,-2.6) node{{\small $(iii)$}};
	\end{tikzpicture}
	\vspace{-0.7cm}
	\caption{(i) In situ density profile after longitudinal expansion from a harmonic trap of a 1d cloud of $N = 4600\pm 100$ $^{87}{\rm Rb}$ atoms ; the smooth curve is the theoretical prediction of GHD and the noisy one is the experimental data. (ii) Initial profile obtained from the Yang-Yang equation of state (YY), Gross-Pitaevskii (GP), ideal Bose gas (IBG) and classical field (CF) \cite{SMalternate}, with the same temperature and chemical potential as for YY. (iii) Evolution from the YY initial profile with GHD and conventional hydrodynamics (CHD).}
	\label{fig:expansion_harm}
	\vspace{-0.5cm}
\end{figure}
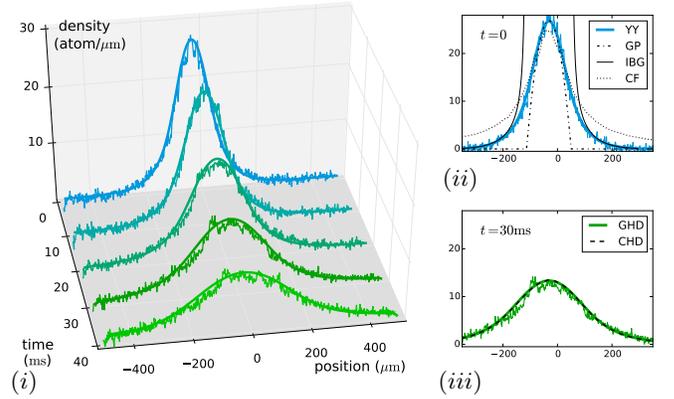

Remarkably, the {\it same} emergent hydrodynamics was rediscovered in 2016 in the 
context of 1d {\it quantum integrable models}  \cite{ghd,bertini1}---the resulting theoretical framework is now dubbed {\it Generalized HydroDynamics} (GHD)---. Cold atom experiments offer a unique platform to test the validity of this theoretical 
breakthrough. Indeed 1d clouds are 
well described by the 1d Bose gas with contact repulsion \cite{Amerongen-Yang-2008,Jacqmin-Sub-Poissonian-2011,Volger-Thermodynamics-2013}, a paradigmatic integrable system known as the Lieb-Liniger model \cite{LiebLiniger} whose large-scale dynamics is argued to be given by GHD \cite{ghd,dy,ddky,doyon_spohn,Berkeley_prl,ghd_qnc,viewpoint,Berkeley_bbh,GHDdiffusion,takato_proof,alvise_lorenzo}. Many other integrable models are argued to be described by GHD, leading to intense research activity in the past two years \cite{bertini1,h1,h2,h3,h4,h5,h6,h7,h8,h9,h10}.

The goal of this Letter is to establish {\it experimentally} GHD as the correct hydrodynamic description of the 1d Bose gas. To do so, we measure the in situ density profiles of a time-evolving 1d atomic cloud trapped on an atom chip, and compare the data with predictions from GHD. We contrast those predictions with the ones of the conventional hydrodynamic (CHD) approach---based on the assumption of local thermal equilibrium \cite{footnote1}---that has been frequently used \cite{chd0,chd1,chd2,chd3,chd4,chd5}. Starting from a cloud at thermal equilibrium in a longitudinal potential $V(x)$, dynamics is triggered by suddenly quenching $V(x)$. We consider three types of quenches. The first is a 1d expansion of the cloud from an initial harmonic potential (Fig. \ref{fig:expansion_harm}); the second is a 1d expansion from a double-well potential (Fig. \ref{fig:dw_expansion}); the third is a quench from double-well to harmonic potential (Fig. \ref{fig:qnc}). We find that only GHD is able to accurately describe the time-evolution of the cloud beyond the harmonic case.

\vspace{0.1cm}
\noindent {\bf\em Generalized HydroDynamics (GHD).}\;The Hamiltonian that describes our atomic gas of $N$ bosons of mass $m$ confined in a potential $V(x)$ with contact repulsion is
\begin{equation}
	\label{eq:LL_ham}
	H  =   -\frac{ \hbar^2  }{2m}  \sum_{i=1}^N \partial_{x_i}^2   +  g\sum_{i < j} \delta(x_i-x_j) + \sum_{i=1}^N V(x_i) ,
\end{equation}
which reduces to the model solved by Lieb and Liniger \cite{LiebLiniger} when $V(x)=0$.
As in any hydrodynamic approach, the idea is to trade that microscopic model for a simpler, long-wavelength, description in terms of continuous densities. The fluid consists 
of local ``fluid cells'' of size $\delta x$, with $\delta x$ very short compared to the wavelength of density variations, but very long compared to microscopic lengths in the gas. 
The state inside each local fluid cell $[x,x+\delta x]$ is a macrostate of the Lieb-Liniger model that is entirely characterized by its distribution $\rho(x,v)$ of rapidities $v$, similarly to the celebrated {\it thermodynamic Bethe Ansatz} of Yang and Yang \cite{yangyang,zamo} (as explained in Refs.~\cite{ghd,bertini1}, that these are the correct local macrostates can be seen as a consequence of recent results on ``generalized thermalization" \cite{mosselcaux,reviewcharges}). Semiclassically, one may view the rapidity $v$ as the bare velocity of a {\it quasi-particle}. As in the 1d billiard, the velocity $v$ gets renormalized in the presence of other quasi-particles, resulting in an effective velocity that is the solution of an integral equation \cite{dobrods,spohn_book,ghd,bertini1},
\begin{subequations}
\label{eq:ghd}
\begin{equation}
	v^{\rm eff} (v) = v + \int d w  \, \rho(w) \Delta (v-w) \left[ v^{\rm eff} (v) - v^{\rm eff} (w)  \right] .
\end{equation}
However, while in the classical billiard the jumping distance $\Delta (v- w)$ at each collision is a constant---the length of the rods---, in the Bose gas it corresponds to the {\it time delay} resulting from the {\it two-body scattering phase} $\phi(v-w)$ through differentiation \cite{LiebLiniger}, $\Delta (v- w) = -\frac{\hbar}{m} \frac{d \phi(v- w)}{d (v-w)} $. This gives $\Delta (v-w) = -\frac{2 g/m}{(g/\hbar)^2 + (v-w)^2}$ for the Dirac delta potential (see Ref. \cite{dyc} for an extended discussion).
The effective velocity enters the evolution equation for the distribution $\rho(x,v)$ as follows \cite{ghd,bertini1,dy}:
\begin{equation}
	\partial_t \rho + \partial_x [v^{\rm eff} \, \rho ] \, = \, \left(\frac{\partial_x V }{m}  \right) \,  \partial_v \rho .
\end{equation}
\end{subequations}
This resembles a Liouville equation for phase space densities of quasi-particles, although it is to be stressed that it is an {\it Euler hydrodynamic equation}, determining the evolution of the degrees of freedom emerging at large wavelengths. {\it Generalized HydroDynamics} consists of Eqs. (\ref{eq:ghd}.a) and (\ref{eq:ghd}.b).
In practice, for a given initial distribution $\rho(x,v)$, the GHD equations can be efficiently solved numerically \cite{ddky,Berkeley_bbh,dyc,ghd_qnc}; in this Letter we rely on a finite-difference method similar to the one discussed in Ref.~\cite{Berkeley_bbh}. Importantly, for our purposes, the atomic density $n(x)$ is obtained from the distribution $\rho(x,v)$ by integrating locally over all rapidities, $n(x) = \int  dv  \, \rho (x, v)$.

\begin{figure}[t]
	\begin{center}
	\includegraphics[width=1.08\linewidth]{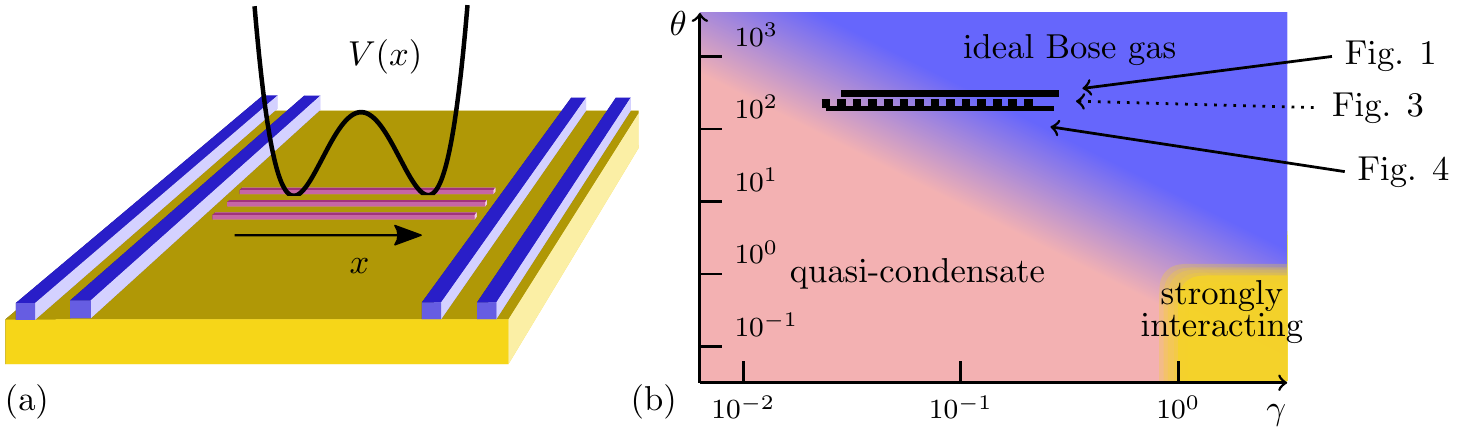}
	\vspace{-0.6cm}
	\caption{(a) The atom chip setup with the four wires (blue) creating the longitudinal  potential and the three wires (red) creating the strong transverse confinement. (b) Position of our three data sets in the thermal equilibrium phase diagram of the Lieb-Liniger gas with $\gamma = mg/(\hbar^2 n)$ and $\theta = 2 \hbar^2 k_{\rm B} T /(m g^2)$~\cite{phasediag}. At the center of the cloud $\gamma$ is of order  $10^{-2}$, but it increases in the wings as the density decreases; we display the segments $[\gamma_{\rm min}, \gamma_{\rm min}/10]$ corresponding to a local density $n(x)$ not smaller than a tenth of the maximal density in the cloud. The asymptotic regimes of the Lieb-Liniger gas, separated by smooth crossovers, are shown in colors. Our data sets lie at the crossover between the quasicondensate and the ideal Bose gas (IBG) regimes.}
	\label{fig:phase_diag}
	\end{center}
	\vspace{-0.5cm}
\end{figure}

\vspace{0.1cm}
\noindent {\bf\em The atom chip.} Our experimental setup 
is described in detail in Ref.~\cite{these-Aisling}. $^{87}$Rb atoms are confined in a magnetic trap produced by microwires deposited on the surface of a chip. The transverse confinement is provided by three \unit[1.3]{mm} long parallel wires (red 
wires in Fig.~\ref{fig:phase_diag}), which carry AC currents modulated at \unit[400]{kHz}: atoms are guided
along $x$, at a distance of \unit[12]{$\mu$m} above the central wire, with a transverse frequency $\omega_\perp$ which lies between 5 and 8 kHz. 
The modulation
technique permits an independent control of the longitudinal potential $V(x)$, which
is realized by two pairs of wires perpendicular to $x$, running DC current (blue wires in Fig.~\ref{fig:phase_diag}). The atomic cloud is far from those wires, in a region where $V(x)$ is well approximated by its Taylor expansion at small $x$. By tuning the currents in the four wires, we effectively control 
the coefficients of the $x$, $x^2$, $x^3$ and $x^4$ terms in that expansion: we can thus produce harmonic potentials, but also double-well potentials.

Using radio-frequency evaporative cooling we produce cold atomic clouds in the 1d regime, with
a typical energy per atom smaller than the transverse energy gap: the temperature and chemical potential fulfill $k_{\rm B} T,\mu <\hbar \omega_\perp$. The gas is then well described by the 1d model (\ref{eq:LL_ham}), with 
the effective 1d repulsion strength  $g = 2 \hbar a \omega_{\perp}$ \cite{olshanii_LL} where the 3d scattering length of $^{87}{\rm Rb}$ is $a = 5.3 \, {\rm nm}$, and the mass is $m = 1.43\times 10^{-25}$kg.
Moreover the lengthscale on which $n(x)$ varies is much larger than microscopic lengths ---the phase correlation length at thermal equilibrium, which is the largest microscopic length in the quasicondensate regime, is of order $n\hbar^2/(m  k_{\rm B}T)$ \cite{Castin-Extension-2003,Cazalilla-Bosonizing-2004}, typically $0.1 \,\mu$m for our clouds--- so the hydrodynamic description applies. At equilibrium, the latter is equivalent to the Local Density Approximation (LDA), 
and the local properties of the gas are parametrized by the dimensionless repulsion strength $\gamma = m g/(\hbar^2 n)$ and the dimensionless temperature $\theta = 2 \hbar^2 k_{\rm B} T/(m g^2)$~\cite{phasediag}. The range $(\gamma, \theta)$ explored by our data sets is displayed in Fig.~\ref{fig:phase_diag}.(b). In this Letter we analyze the density profiles $n(x)$, which we measure using absorption images~\cite{these-Aisling}, averaging over typically ten images, with a pixel size of $1.74~\mu$m in the atomic plane.

\vspace{0.1cm}
\noindent {\bf\em The Yang-Yang initial profile.}\quad We start by trapping a cloud of $N =4600\pm 100$ atoms, with $\omega_\perp = 2\pi \times (7.75 \pm 0.02) \, {\rm kHz}$, in a harmonic potential $V(x)=m \omega_\parallel^2 x^2/2$ with $\omega_\parallel = 2\pi \times (8.8 \pm 0.04 )\,{\rm Hz}$, and measure its density profile (Fig.~\ref{fig:expansion_harm}.(ii)). To evaluate the temperature of the cloud, we fit the experimental profile with the one predicted by the Yang-Yang equation of state \cite{yangyang,Amerongen-Yang-2008,Jacqmin-Sub-Poissonian-2011,Volger-Thermodynamics-2013}, relying on LDA and on the assumption that the cloud is at thermal equilibrium; we find $T  = (0.43 \pm 0.013)\, \mu {\rm K}$. This gives $\theta = (3.5 \pm 0.1) \times 10^2$, while the interaction parameter is $\gamma =(2.8 \pm 0.1) \times 10^{-2}$ at the center.

As the density varies from the center of the cloud to the wings, the gas locally explores several regimes \cite{phasediag}, from quasicondensate to highly degenerate Ideal Bose Gas (IBG) to non-degenerate IBG, see Fig.~\ref{fig:phase_diag}(b). The Yang-Yang equation of state \cite{yangyang} is exact in the entire phase diagram of the Lieb-Liniger model, and thus faithfully describes the density profile within LDA. We stress that this is the most natural and powerful method to describe the initial state of the gas \cite{Amerongen-Yang-2008,Jacqmin-Sub-Poissonian-2011,Volger-Thermodynamics-2013}, and that no simpler approximate theory \cite{SMalternate} can account for the whole initial density profile, see Fig.~\ref{fig:expansion_harm}(ii). The Gross-Pitaevskii (GP) theory works in the central part---because it is close to the quasicondensate regime---, but not in the wings. The opposite is true for the IBG model: it correctly describes the wings, but not the center of the cloud---the chemical potential is positive in the center, so the density diverges in the IBG---. The classical field theory captures the quasicondensation
transition for gases deep in the weakly interacting regime  but it fails to reproduce faithfully the wings of our cloud since the latter are not highly degenerate.

\vspace{0.1cm}
\noindent {\bf\em Expansion from harmonic trap: agreement with both GHD and CHD.}\; At $t=0$, we suddenly switch off the longitudinal harmonic potential $V(x)$, and let the cloud expand freely in 1d. We measure the in situ profiles at times $t=10,20,30$ and 40 ms, see Fig.~\ref{fig:expansion_harm}(i).

Two theories are able to give predictions for the expansion starting from the locally thermal initial state.
One is GHD, presented above, where the full distribution of quasi-particles $\rho(x,v)$ is evolved in time~\cite{footnote3}. The other is the {\it conventional} hydrodynamics (CHD) of the gas which, contrary to GHD, assumes that all local fluid cells are at thermal equilibrium, and keeps track only of three quantities that entirely describe the local state of the gas: the density $n(x)$, the fluid velocity $u(x)$, and the internal energy $e(x)$ \cite{SMalternate}. We calculate the evolution of the density profile with both theories, and find that both of them are in excellent agreement with the experimental data, see Fig.~\ref{fig:expansion_harm}(iii) for the result at $t=30$~ms.

GHD and CHD thus appear to be indistinguishable in that situation, at least for the expansion times that we probe here. We attribute this coincidence to the initial {\it harmonic} potential, which is very special. In this case it is simple to see that the GHD and CHD predictions coincide in the ideal Bose gas regime, and they can be shown to stay relatively near even beyond that regime \cite{SMharmonic}.
\begin{figure}[t]	
	\begin{tikzpicture}
		\draw (0,0) node{\includegraphics[width=0.32\textwidth]{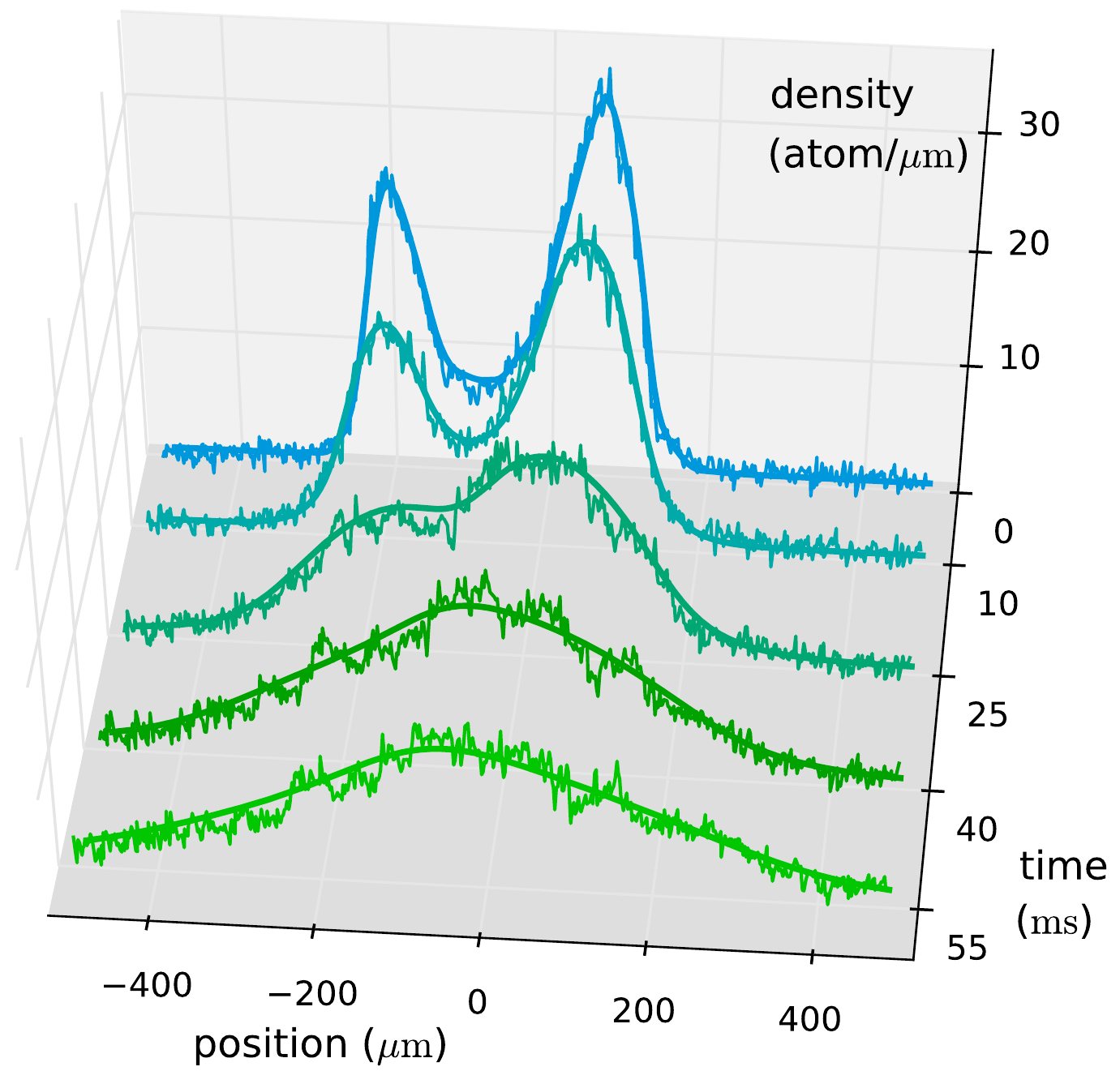}};
		\draw (-2.6,-2.6) node{$(i)$};
		\draw (3.2,-2.6) node{$(ii)$};
		\draw (4.3,-1.5) node{\includegraphics[width=0.155\textwidth]{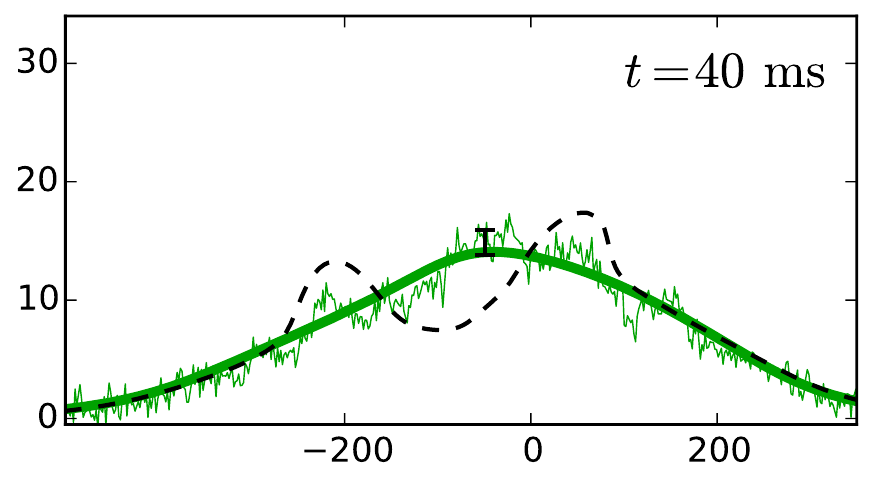}};
		\draw (4.3,0.1) node{\includegraphics[width=0.155\textwidth]{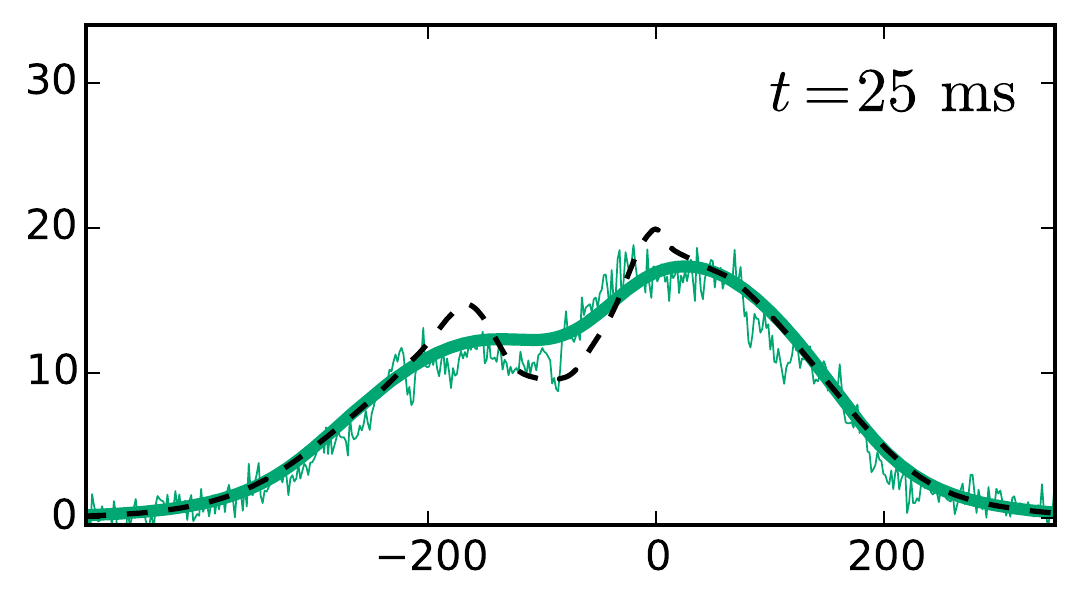}};
		\draw (4.3,1.7) node{\includegraphics[width=0.155\textwidth]{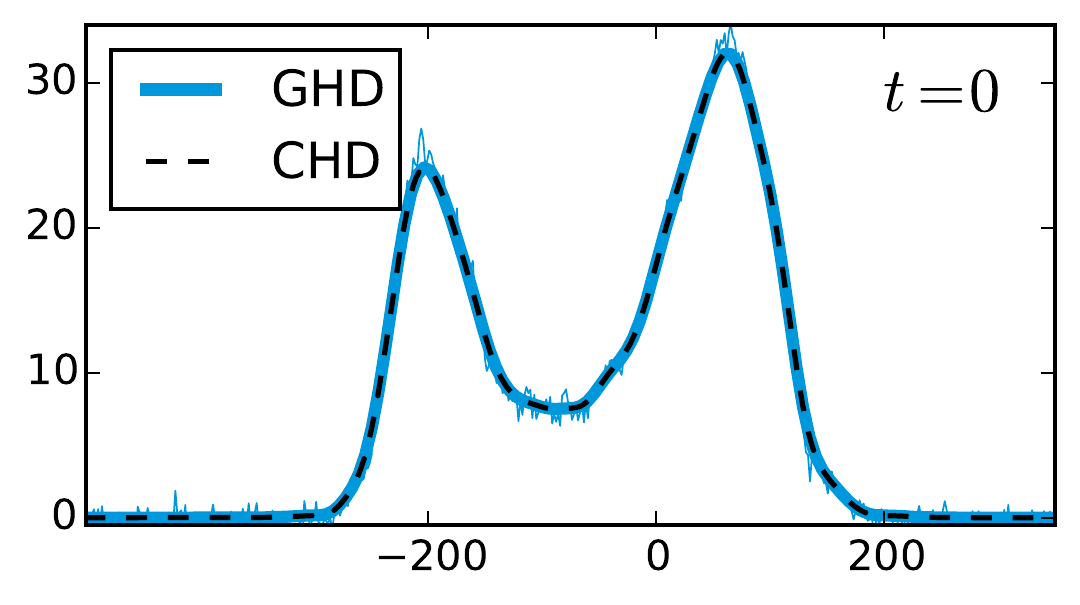}};
	\end{tikzpicture} \\
	\vspace{-0.3cm}
	\caption{(i) Longitudinal expansion of a cloud of $N = 6300\pm 200$ atoms initially trapped in a double-well potential, compared with GHD. (ii) Even though the initial state is the same for GHD and CHD, both theories clearly differ at later times. CHD wrongly predicts the formation of two large density waves. The error bar shown at the center at $t=40$ms corresponds to a 68\% confidence interval, and is representative for all data sets.}
	\label{fig:dw_expansion}
	\vspace{-0.3cm}
\end{figure}

\vspace{0.1cm}
\noindent {\bf\em Discussion: GHD vs. CHD.}\; We wish to identify a setup where the theoretical predictions of both theories clearly differ, in order to  experimentally discriminate between them. This will be the case if GHD predicts, for  some time $t$ and at some position $x$, that the distribution of rapidities $\rho(x,v)$ will differ strongly from a thermal equilibrium one.

Such a situation occurs during the expansion of a cloud that initially has two well separated density peaks (Fig.~\ref{fig:dw_expansion}). The reason can be captured by the following argument. The fluid cells $[x, x+\delta x]$ that are around either of the two peaks contain more quasi-particles, including quasi-particles of large rapidities, than the fluid cells near the center at $x=0$.
Under time-evolution, the quasi-particles from the left peak that have a large positive rapidity $+u$ soon meet the ones coming from the right peak that have a large negative rapidity $-u$,  around $x=0$. Then,
the distribution of rapidities near $x=0$ is double-peaked, with maxima at $v\simeq \pm u$, so it is clearly very far from a thermal equilibrium distribution, which would be single-peaked. This phenomenon is obvious for non-interacting particles, Eq.~(\ref{eq:ghd}) reducing to the standard Liouville equation, and GHD calculations indicate that this is true also for interacting particles \cite{ghd_qnc,SMphasespace}.

\vspace{0.1cm}
\noindent {\bf\em Expansion from a double-well.}\; To realize the above scenario, we prepare a cloud of $N = 6300\pm 200$ atoms, with $\omega_\perp = 2\pi \times (8.1 \pm 0.03)\,{\rm kHz}$, at thermal equilibrium in a longitudinal double-well potential $V(x)$, such that the atomic density presents two well separated peaks, the
peak density corresponding to $\gamma = (2.45 \pm 0.07) \times 10^{-2}$. Then at $t=0$ we suddenly switch off the potential $V(x)$ and measure the in situ profiles at time $t=10$, $25$, $40$, $55 \, {\rm ms}$ (Fig. \ref{fig:dw_expansion}).

To compare with theoretical predictions, we need to know the initial temperature $T$ of the cloud. However we cannot estimate $T$ from fitting 
the initial density profile $n_0(x)$ with the Yang-Yang equation of state and LDA because we do not have a good knowledge
of the initial potential $V(x)$ that we create on the chip. Instead, we proceed as follows. First we postulate an initial
temperature $T$ and construct the
initial rapidity distribution $\rho_T(x,v)$ such that, for a given $x$, 
$\rho_T(x,v)$ is the
thermal equilibrium rapidity distribution of Yang-Yang~\cite{yangyang} at temperature $T$
and density $n_0(x)$. 
We then evolve $\rho_T(x,v)$ using GHD and compute $n_T(x,t)$.
While, by construction,  $n_T(x,0)=n_0(x)$,
$n_T(x,t)$ may differ from the data at later times.
We repeat this procedure for
several initial temperatures and we select the value of $T$
whose time evolution is in best  agreement
with the data \cite{SMprocedure}. We obtain $T \simeq 0.3 \, \mu {\rm K}$,
corresponding to  $\theta \simeq 2 \times 10^2$, see Fig. \ref{fig:phase_diag}(b).

The comparison between the expansion data and GHD is shown in Fig. \ref{fig:dw_expansion}(i); the agreement is excellent. We also simulate the time-evolution of the cloud with CHD, for the exact same initial state. As we expected, expanding from a double-well potential reveals a clear difference between CHD and GHD, see Fig.~\ref{fig:dw_expansion}(ii). Two large density waves emerge in CHD and large gradients develop, eventually leading to shocks~\cite{ddky}, features which are not seen in GHD~\cite{SMphasespace}. 

\begin{figure}[t]
	\hspace{-0.4cm}
	\includegraphics[width=0.5\textwidth]{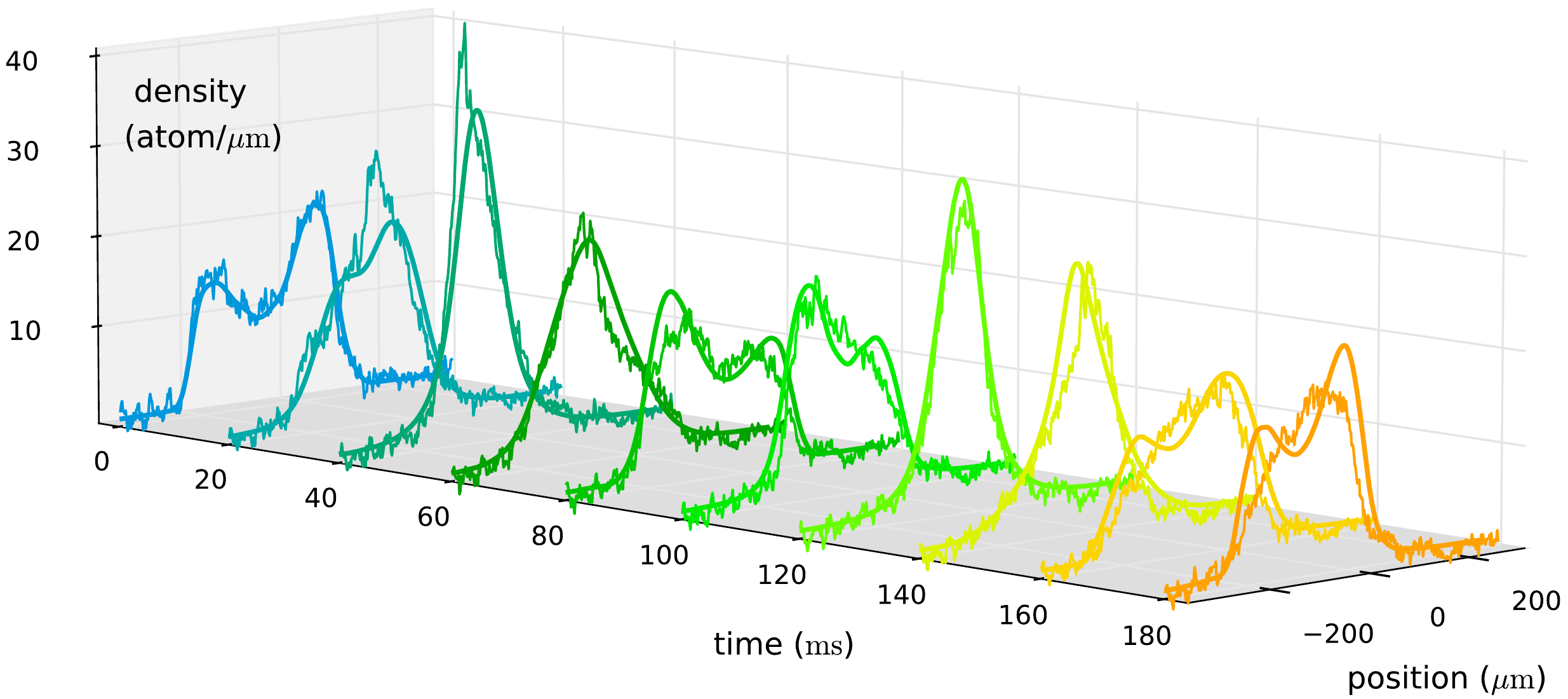} \\
	\vspace{-0.3cm}
	\caption{Quench from double-well to harmonic potential, compared to the GHD prediction, with an atomic cloud that contains $N = 3500 \pm 140$ atoms initially. The main features of the experimental data are well reproduced by GHD. One experimental effect, not modeled in GHD, that appears to be particularly important, are the three-body losses: after 180~ms, the number of atoms drops by approximately $15 \%$.}
	\label{fig:qnc}
	\vspace{-0.3cm}
\end{figure}

\vspace{0.1cm}
\noindent {\bf\em Quench from double-well to harmonic potential.}\;Finally, we trap $N = 3500 \pm 140$ atoms, with $\omega_\perp = 2\pi \times (5.4 \pm 0.02) \,{\rm kHz}$, in a double-well potential, and we study the evolution of the cloud after suddenly switching off the double-well and replacing it by a harmonic potential of frequency $\omega_\parallel = 2\pi \times (6.5 \pm 0.03) \,{\rm Hz}$. We measure the in situ profiles at time $t=0,20,40,\dots, 180$~ms, see Fig.~\ref{fig:qnc}. The initial peak density corresponds to $\gamma  = (2.13 \pm 0.07 ) \times 10^{-2}$. To estimate the temperature of the cloud, we proceed as in the previous case \cite{SMprocedure}; we find $T \simeq 0.15 \, \mu {\rm K}$, corresponding to $\theta \simeq 2.2\times 10^2$ (Fig. \ref{fig:phase_diag}(b)).

This quench protocol mimics the famous quantum Newton's Cradle experiment \cite{qnc} ---see also Refs.~\cite{qnc2,qnc3} for recent realizations---, which is realized here in a weakly interacting gas. Exactly like in the previous paragraph, this is a situation where GHD predicts the appearance of non-thermal rapidity distributions \cite{ghd_qnc,SMperiodicity}, and must therefore differ strongly from CHD. In fact, we have observed that CHD develops a shock at short times (around $t \simeq 30$~ms), so it is simply unable to give any prediction for the whole evolution time investigated experimentally~\cite{footnote2}.

Importantly, the motion is not periodic, contrary to what would be seen purely in the IBG or in the strongly interacting fermionized regime. Nevertheless, the motion of the cloud preserves an {\it approximate} periodicity, with a period close to, but slightly longer than, $2\pi/\omega_\parallel$~\cite{SMperiodicity} (of course, if the cloud was symmetric under $x \rightarrow -x$, the period would be divided by two).
At a quarter of the period ---and three quarters of the period---, the density 
distribution shows a single thin peak located near $x=0$. We find good agreement with the GHD predictions,
with the initial temperature $T$ as the only free parameter~\cite{SMprocedure}. However, experimental effects not taken into account by the GHD equations~(\ref{eq:ghd}) appear to be more important in this setup than in the previous ones of Figs.~\ref{fig:expansion_harm}-\ref{fig:dw_expansion}, where shorter times were probed. For instance, the number of atoms $N$ is not constant in our experimental setup: it decreases with time and drops by approximately $15 \%$ after $180$~ms, probably because of three-body losses which occur at large density. This might partially explain the difference between the experimental 
density profile and the GHD one. We also suspect the small residual roughness of the  potential $V(x)$ of affecting the experimental profiles.

\vspace{0.1cm}
\noindent {\bf\em Conclusion.}\quad The results presented in this Letter are the first experimental check of the validity of GHD for 1d integrable quantum systems. We have shown  that GHD ---which predicts the time evolution of the distribution of rapidities--- accurately captures the motion of 1d cold bosonic clouds made of $N \sim 10^3$ atoms, on time scales of up to $\sim 0.2$~seconds. We probed situations where the GHD predictions significantly differ from the ones of the conventional hydrodynamic approach, even at short times. We stress that GHD is applicable to all regimes of the 1d Bose gas, and it would therefore be particularly interesting to probe the strongly interacting regime. More generally, GHD is applicable to all Bethe Ansatz solvable models, including multicomponent mixtures of fermions and bosons with symmetric interactions~\cite{Yang1967,sutherland,gaudin,batchelor}, so it would be very exciting to use it to describe the dynamics of more complex gases which can be realized in experimental setups different from ours~\cite{whitlock,fallani}.



\begin{acknowledgments}
We thank R. Dubessy for very stimulating discussions, and M. Cheneau for his careful reading of the manuscript. BD and JD thank J.-S. Caux, R. Konik and T. Yoshimura for joint work on closely related projects. MS and IB thank S. Bouchoule of C2N (Centre Nanosciences et Nanotechnologies, CNRS/UPSUD, Marcoussis, France) for the development and microfabrication of the atom chip. A. Durnez and A. Harouri of C2N are acknowledged for their technical support. C2N laboratory is a member of RENATECH, the French national network of large facilities for micronanotechnology. 

MS acknowledges support by the Studienstiftung des Deutschen Volkes. This work was supported by R\'egion \^{I}le de France (DIM NanoK, Atocirc project), and by the CNRS Mission Interdisciplinaire ``D\'efi Infiniti'' (JD). BD thanks the Centre for Non-Equilibrium Sciences (CNES) and the Thomas Young Centre (TYC), as well as the Perimeter Institute (Waterloo, Canada) for hospitality while some of this work was done (September 2018); JD also thanks SISSA in Trieste for hospitality. BD and JD thank the Galileo Galilei Institute for Theoretical Physics (Florence, Italy) for hospitality during the workshop ``Entanglement in Quantum Systems'' (June 2018), and the Erwin Schroedinger Institute (Vienna, Austria) for hospitaliry during the programme ``Quantum Paths" (April-May 2018).
\end{acknowledgments}

\vspace{2cm}

\pagebreak
\newpage

\begin{widetext}

\begin{center}
{\bf Supplemental Material for ``Generalized Hydrodynamics on an Atom Chip''}
\end{center}

\section{Regimes of the Lieb-Liniger gas at thermal equilibrium}

Properties of the Lieb-Liniger gas at thermal equilibrium have been studied extensively. The gas at thermal equilibrium is entirely parametrized by the dimensionless parameters $\gamma$ and $\theta$. 
Three main regimes have been identified, which correspond to 
different asymptotic value of the normalized zero-distance two-body correlation
function $g^{(2)}(0)$~\cite{phasediag}. The region in the ($\gamma,\theta$) plane corresponding to $\gamma \ll 1$ and $\theta \ll \gamma^{-3/2}$ is that of the quasicondensate regime. In this regime, correlations between particles 
are small and $g^{(2)}(0)\simeq 1$. Thus locally, the gas ressembles a 
Bose-Einstein condensate, although quantum and thermal fluctuations
associated to long wavelength phonons prevent true long range order.
The region $\theta \gg \rm{max}(1, \gamma^{-3/2})$ is that of the 
ideal Bose gas regime, characterized by $g^{(2)}(0)\simeq 2$. 
In this regime repulsive
interactions are not strong enough to prevent the thermally activated 
large density fluctuations associated to the bosonic bunching phenomenon.
Finally, the region $\theta\ll 1$ and $\gamma\gg 1$ is that of 
the strongly interacting regime, also called Tonks-Girardeau or fermionized regime. There, interactions are strong enough to almost prevent two atoms to sit at the same position so that $g^{(2)}(0)\ll 1 $.

The above-mentioned regimes are all separated by smooth crossovers. 
Quantitative values for the position of these 
transitions depend on the quantity that is considered and on the 
criteria used. One possible criterion to locate the 
transition from the IBG to qBEC regime is the condition $\mu=0$, 
where $\mu$ is the chemical potential: in the IBG regime, $\mu<0$ 
while $\mu>0$ in the qBEC regime ($\mu\simeq gn$ in the qBEC regime). 
In Fig.~\ref{fig:app:phasediag}, we plot the line $\mu=0$, found using the Yang-Yang equation of state. This line follows 
the expected scaling $\theta \propto \gamma^{-3/2}$ of the crossover 
between the IBG and the qBEC regime. It lies however substantially above 
the line  $\theta = \gamma^{-3/2}$ \cite{phasediag}.
 
The IBG regime can be divided into two subregimes: the highly 
non-degenerate regime corresponding to $\theta \gg \gamma^{-1/2}$ and 
the highly degenerate regime corresponding to 
$\gamma^{-3/2}\ll \theta \gg \gamma^{-1/2}$. The crossover between these two 
subregimes is very wide and in fig.~\ref{fig:app:phasediag}, we signal this transition
with  the line corresponding
to $n\lambda_{dB}=1$, where $\lambda_{dB}=\hbar/\sqrt{2\pi m k_B T}$ is the de Broglie 
wavelength.

\begin{figure}\centerline{\includegraphics{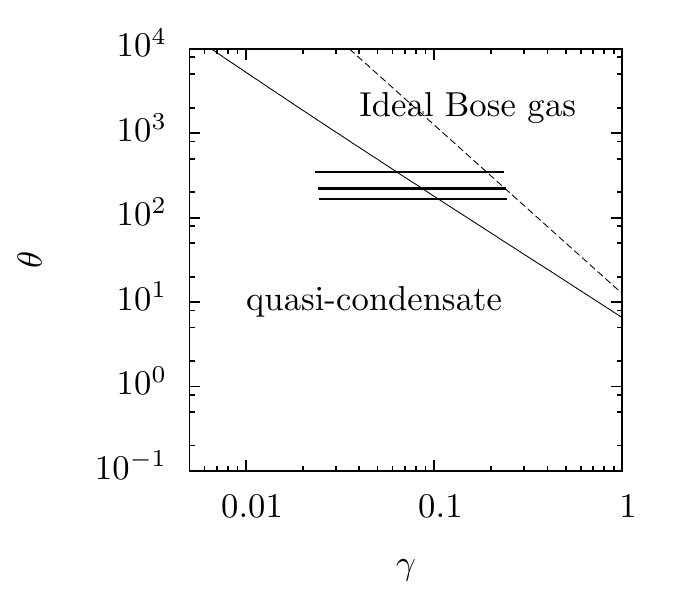}}
\caption{Position of the three sets of data shown in the text in the $(\gamma,\theta)$ space. The thin solid line is the line corresponding to a vanishing chemical potential, computed using Yang-Yang equation of state, which indicates the crossover from the Ideal Bose gas to the quasi-condensate. The dashed line is the line $n\lambda_{\rm{dB}}=1$ which indicates the crossover from a non-degenerate to a highly degenerate gas. The horizontal fat lines correspond to the experimental data investigated in the main text  (from top to bottom data of Fig.1, data of Fig.3 and data of Fig. 4). }
\label{fig:app:phasediag}
\end{figure}

\section{Approximate theories that we compare to GHD}

\subsection{Conventional hydrodynamics based on Yang-Yang thermodynamics (or Thermodynamic Bethe Ansatz)}

What we call `conventional' hydrodynamics (CHD) in the main text is simply the Euler equations that hold in normal fluids that are locally at thermal equilibrium, that govern the variation of the particle density $n(x,t)$, the fluid velocity $u(x,t)$, and the internal energy per particle $e(x,t)$:
\begin{equation}
 	\left\{ \begin{array}{rcl}
	 	\partial_t n + \partial_x (u n) & = & 0 \\
	 	\partial_t u + u \partial_x u   + \frac{1}{m n} \partial_x P  & = &  - \frac{1}{m} \partial_x   V   \\
	 	\partial_t e + u \partial_x e   + \frac{P}{n} \partial_x u  & = & 0   .
 	\end{array} \right.
 \end{equation}
Here $P(n,e)$ is the pressure, which is given by the equation of state of the fluid. One can put these three equations in a form that expresses conservation of mass, momentum---broken by the external potential $V(x)$---, and energy,
\begin{equation}
 	\left\{  \begin{array}{rcl}
	 	\partial_t n + \partial_x (u n) & = & 0 \\
	 	\partial_t (m n u) +  \partial_x (u(m n u)  +P)  & = &   - n \partial_x   V   \\
		\partial_t (n \frac{m u^2}{2} + n e + n V) +  \partial_x ( u (n \frac{m u^2}{2} + n e + n V  ) + u P)   & = & 0  . 
 	\end{array} \right.
 \end{equation}
These are the evolution equations we solve to obtain the Conventional HydroDynamics (CHD) curves in the main text. The pressure $P(n,e)$ is obtained from the thermodynamic Bethe Ansatz, also known as the Yang-Yang equation of state \cite{yangyang}.

One convenient way to calculate the pressure is to observe that it is the momentum current, and then to use the formula from \cite{ghd} for the latter,
\begin{equation}
	\label{eq:pressure}
	P \, = \,  \frac{m^2}{\hbar} \int \frac{d v}{2\pi} v \, \nu (v)  \, v^{\rm dr} .
\end{equation}
Here $\nu(v) \in [0,1]$ is the occupation function, which, for the thermal Gibbs state at inverse temperature $\beta = 1/(k_{\rm B} T)$, satisfies the integral equation of Yang and Yang \cite{yangyang},
\begin{equation}
	\label{eq:TBA}
	\beta \frac{m v^2}{2} \, = \, \log \left[ \frac{1-\nu(v)}{\nu (v)} \right] \, - \, \int \frac{d w}{2\pi} K(v-w) \log\left[ 1-\nu (w) \right] ,
\end{equation}
where $K(v - w) = \frac{2 (v-w)/\hbar}{(g/\hbar)^2 + (v-w)^2}$ is the differential scattering phase, or Lieb-Liniger kernel. The superscript 'dr' in (\ref{eq:pressure}) stands for 'dressing', and is defined as follows for a function $f(v)$,
\begin{equation}\label{eq:fdr}
	f^{\rm dr} (v) \, = \, f(v) + \int \frac{d w}{2\pi  } K(v-w) \nu (w) f^{{\rm dr}}(w) 
\end{equation}
(and by a slight abuse of notation, we write $v^{\rm dr}$ for ${\rm id}^{\rm dr}(v)$ where ${\rm id}(v) = v$ is the identity function). The occupation function $\nu(v)$ and the distribution $\rho(v)$ used in the main text are related by
\begin{equation}
	\label{eq:nu_rho}
    \rho(v) = \frac{m}{2\pi \hbar} \, \nu(v) \, 1^{\rm dr} (v)
\end{equation}
where $1(v)=1$. The function $\nu(v)$ gives the occupation fraction of the rapidities, which 
are the fermionic quantities defining a Bethe-Anstaz eigenstate~\cite{yangyang}.

\subsection{Gross-Pitaevskii approach}
The Gross-Pitaevskii description (GP) assumes that the gas is Bose-condensed : all the atoms are in the same single-particle wave function $\psi(x)$. It describes a Bose gas at zero temperature in the limit of weak repulsion $\gamma \rightarrow 0$, as long as their extension is much smaller
than $\xi e^{\pi/\sqrt{\gamma}}$, where $\xi=\hbar/\sqrt{m\rho}$ is the healing length. The latter 
condition, usually fulfilled in experiment with weakly interacting gases, ensures that quantum fluctuations do not break long-range order.
At equilibrium the wavefunction $\psi$ obeys 
\begin{equation}
\mu\psi=\, \left(-\frac{\hbar^2}{2m} \partial_x^2 + g |\psi|^2 + V(x)  \right) \psi
\end{equation} 
where $\mu$ is the chemical potential.
After modification of the potential, the time evolution of $\psi$ is given by the Gross-Pitaevskii equation 
\begin{equation}
	i \hbar \partial_t \psi \, = \, \left(-\frac{\hbar^2}{2m} \partial_x^2 + g |\psi|^2 + V(x)  \right) \psi .
\label{eq.GP}
\end{equation}
Here $\psi (x, t)$ is the wavefunction of the condensate at time $t$, normalized such that
\begin{equation}
	 \int dx |\psi(x,t)|^2 \, = \, N .
\end{equation}
Separating the phase and the amplitude of the wavefunction of the condensate,
 \begin{equation}
 	\psi(x,t) \, = \, \sqrt{n(x,t)} e^{i \varphi(x,t)} ,
 \end{equation}
and introducing the velocity field  $u = \frac{\hbar}{m} \partial_x \varphi$, 
one gets GP in Madelung form,
 \begin{equation}
 	\begin{array}{rcl}
	 	\partial_t n + \partial_x (u n) & = & 0 \\
	 	\partial_t u + u \partial_x u & = &  - \frac{1}{m} \partial_x  \left[  gn + V(x)  -  \frac{\hbar^2}{2m} \frac{\partial_x^2 \sqrt{n}}{\sqrt{n}}  \right]		.
 	\end{array}
    \label{eq.GPmadelung}
 \end{equation}
 The first equation expresses conservation of mass, the second is the Euler equation for a fluid with the equation of state $\mu = g n$ in an external potential $V(x)$, with an additional {\it quantum pressure} term $\frac{\hbar^2}{2m}   \frac{\partial_x^2 \sqrt{n}}{\sqrt{n}}$. In the {\it Euler limit} of long wavelengths for density variations, the quantum pressure term can be neglected. Then Eq. (\ref{eq.GPmadelung}) reduces to the Euler equation with a pressure term $P=gn^2/2$, which is indeed the pressure of a condensate or a quasicondensate. Note however that the GP equation goes beyond this {\it Euler limit} of long wavelength variation of $\psi$. For instance, it would capture correctly the ``quantum Newton craddle'' experiment presented in the main text, for a weakly interacting cloud at $T=0$. At the ``collision-time'' where the two initial clouds are on top of each other, GP predicts fast oscillations of the density reflecting interference phenomena. 
Of course such fast oscillations, which, in terms of Bethe-Ansatz states, would imply interference between different states (of very similar 
quasi-momentum distribution), is not accounted for by GHD. However we 
expect that predictions of GHD (for a weakly interacting gas at $T=0$)
 would coincide with the GP ones, if one
perform a spatial coarse-grained approximation of the GP solution, 
smearing out the interference fringes.   
It is an open question under which precise conditions and coarse-graining procedure  
the predictions of GHD coincide with the coarse-grained GP predictions. \vspace{0.4cm}


\subsection{The classical field approximation}

In the domain characterized by $ \gamma^{-1}\ll \theta \ll \gamma^{-2}$, the
typical occupation number of the relevant modes (either the one-particle 
states for $\theta \gg \gamma^{-3/2}$, or the Bogolioubov  collective modes 
for $\theta \ll \gamma^{-3/2}$) is much larger than one: in particular the 
correlation functions are mainly dominated by the contribution of 
highly populated modes. In such a condition, a theory that neglects the 
quantification of the atomic field $\hat\psi$ and treat it as a 
complex classical field ---or equivalently, that considers statistical fluctuation of the condensate's single-particle wave function--- is expected to provide the relevant physics. 
This is the so-called classical field approximation \cite{Castin-coherence-2000}.
At thermal equilibrium  the complex field $\psi(x)$ obeys the probability 
distribution
\begin{equation}
P(\{\psi(x)\})=Z^{-1} e^{-\beta\int dx (\hbar^2/(2m)|\partial_x\psi|^2 + g/2 |\psi|^4+(V(x)-\mu)|\psi|^2)},
\label{eq.samplingCF}
\end{equation}  
where $\beta=1/(k_B T)$, $\mu$ is the chemical potential, and the partition 
function $Z$ ensures the correct normalization. At sufficiently shallow confining potential, a local density approximation is valid and the external potential $V(x)$ is not very relevant. Note that while two parameters 
are required to describe a Lieb-Liniger gas,  only one 
parameter, which can be taken as $\eta=\mu(\hbar^2/(mg^2 k_B^2T^2))^{1/3}$ in terms of the
temperature and chemical potential, is sufficient to characterize the classical field, providing lengths and densities are correctly rescaled \cite{Castin-coherence-2000,isabelle_twobody}. 
One way of sampling the field $\psi$ according to Eq.~(\ref{eq.samplingCF})  is 
to numerically evolve $\psi$ under a stochastic differential equation. 

In contrast to the GP approach above, the classical field takes into 
account thermal fluctuations in the system. In particular it captures the 
quasicondensation crossover, which occurs, for $\gamma\ll 1$, 
around the line $\theta=\gamma^{-3/2}$, although very large values of $\theta$ are required for an accurate description of the gas~\cite{jacqmin-momentum}. On the other hand, it ignores the quantification
of atoms. This leads to an overestimation of the population in high energy modes:
within CF theory the mode population is $k_B T/\epsilon_\nu$, regardless of 
the mode energy $\epsilon_\nu$, a value much larger, for modes of 
energy $\epsilon_k \gg k_B T$, than the expected
 exponential behavior $e^{-\epsilon_k/(k_B T)}$ (this is a similar phenomenon as that of the well-known UV catastrophe of thermal ensembles of electromagnetic fields). This explains why the equilibrium profile predicted by the classical field 
model strongly overestimate the wings of the atomic clouds, see Fig.~1(ii) in the main text. 
We are aware of refined theories, based on classical field, that include
a well-chosen cutoff above which the excitations are treated as 
a quantum gas, see \cite{Cockburn-comparison-2011,Blakie} 
and references therein. They permit to reproduce the 
equilibrium density profiles of trapped weakly interacting gases \cite{Cockburn-comparison-2011}.
However, expanding such theories to time-dependent 
problems is a priori not trivial. We also think that, in the same way 
as Yang-Yang thermodynamics offers a much more powerful, reliable and
elegant way of describing the equilibrium profiles \cite{Amerongen-Yang-2008,Jacqmin-Sub-Poissonian-2011,Volger-Thermodynamics-2013}, GHD would
supersede complicated and ad-hoc 
theories based on the classical field approximation.


\section{Relation between GHD and CHD for an expansion from a harmonic potential}

As we observed in the main text, the one-dimensional expansion of the gas starting from a state where it is confined by harmonic potential is well described both by GHD and CHD. This is surprising, as GHD takes into account infinitely many conservation laws, while CHD only takes into account three: mass, momentum and energy. Of course, the initial state is, in both cases, obtained by a Gibbs local density approximation, according to which in every mesoscopic cell the gas is a Gibbs state, thermalized with respect to the energy and number of particles (with zero average momentum). Nevertheless, one generically expects that under the GHD evolution, other conservation laws become involved, the fluid being, after some time, locally a generalized Gibbs ensemble -- involving higher conserved charges -- instead of a Gibbs ensemble.

There are two limits where one can show that the GHD evolution is, in fact, the same as the CHD evolution.

The first is the zero-temperature limit. It was shown in Ref. \cite{ddky} that if the initial state is at zero temperature, where, in every cell $x$, the quasi-particle occupation function $\nu(x,v)$ takes the value 1 on an interval $v\in[v_l(x),v_r(x)]$ and 0 elsewhere, then, at least for small enough times, a GHD evolution is completely equivalent to a CHD evolution, and the state stays at zero temperature. This holds from any initial potential, harmonic or not. The equivalence between the evolutions breaks down when in CHD shocks develop; in GHD, large gradients are temporary, and the fluid passes to a higher-dimensional space of states afforded by the higher conservation laws.

The other limit is that of the free gases - either the ideal Fermi gas (at strong coupling) or the ideal Bose gas (IBG) (at large temperatures). Let us consider the latter, as it is more relevant to the present experiment. In the free limit, the effective velocity becomes equal to the velocity, and the initial fluid's Bosonic occupation function $\tilde\nu(x,v)$ (note that this is related to, but not the same as, the quasi-particle -- or fermionic -- occupation function $\nu(x,v)$) is of the form
\beq
	\t\nu(x,v) = 
    \frc1{e^{(mv^2/2 + m\omega_\parallel^2 x^2/2 - \mu)/T}-1}.
\eeq
The GHD evolution during the expansion is very simple,
\beq
	\p_t \t\nu + v \p_x \t\nu = 0,
\eeq
and thus the solution is obtained as
\beq
	\t\nu(x,t,v) = 
    \frc1{e^{(mv^2/2 + m\omega_\parallel^2 (x-vt)^2/2 - \mu)/T}-1}=
\frc1{e^{(m(v-u(x,t))^2/2)(1+\omega_\parallel^2t^2)/T}e^{(m\omega_\parallel^2/(1+\omega_\parallel^2t^2)\, x^2/2 - \mu)/T}-1},    
\label{eq.expansionharmgazid}
\eeq
where $u(x,t)=\omega_\parallel^2/(1+\omega_\parallel^2t^2)\,xt$.
Eq.~(\ref{eq.expansionharmgazid}) shows that the distribution of $v$ at a given $x$ is that of 
a Gibbs ensemble for a gas whose center of mass moves at velocity $u(x,t)$. Hence, all currents obtained after GHD evolution are evaluated within Gibbs states, completely determined by the first three conserved densities. Therefore, it is sufficient to restrict to the evolution described by the first three conservation laws, and we recover CHD.

These two limits, however, only partially explain the observation we have made: the parameters of the experiment, the full range of regimes covered by the gas (see Fig. 2b in the main text), imply that there are regions where the occupation function is both relatively far from the zero-temperature form (see typical initial quasi-particle occupation functions in section \ref{sectII}) and from the IBG form.

We provide below a calculation that shows that, more generally, one may expect the GHD solution to stay relatively near to CHD solution in an expansion from a harmonic potential, even away from the two limits described above. The main result is that under GHD evolution from a harmonic potential, the occupation function has {\em the same asymptotics at large $v$}, of the form $e^{-P(v)}$ with $P(v)$ a polynomial of degree 2, as that obtained from a CHD evolution. That is, the ``wings" in quasi-momentum space are not drastically affected by the presence of higher conservation laws. Further, at finite $v$, the discrepancies must stay bounded. If the occupation function is near enough to the zero-temperature form, with a region relatively near to 1 at small velocities, surrounded by  wings where it tends towards 0, then a combination of the zero-temperature result recalled above, and the asymptotic results derived below, suggests that indeed the GHD and CHD solutions should stay near. Typical initial occupation functions shown in section \ref{sectII} indeed have this form, and thus these arguments explain our observation.

In terms of the quasi-particle occupation function $\nu(v)$, related to the quasi-particle density $\rho(x,v)$ via \eqref{eq:nu_rho}, it turns out that Eq. (2b) in the main text becomes
\beq
	\label{eq:GHD_nu}
	\p_t \nu + v^{\rm eff} \p_x \nu = 0.
\eeq
The pseudoenergy $\ep$, defined via  $\nu=1/(1+e^\ep)$, hence satisfies the same equation,
\beq\label{ep}
	\p_t \ep + v^{\rm eff} \p_x \ep = 0.
\eeq
In order to describe the macrostate, it is convenient to use the energy function $e(v)$, in terms of which the pseudoenergy can be expressed as the solution to the non-linear integral equation \cite{yangyang}
\beq\label{w}
	\ep(v) = e(v) - \int \frc{d w}{2\pi}\,K(v-w) \log (1+e^{-\ep(w)}).
\eeq
Like the quasi-particle density, the energy function $e(v)$, in GHD, becomes space-time dependent, and we can re-write \eqref{ep} using \eqref{w}. One can show that \cite{dy}
\beq\label{wdr}
	(\p_t e)^{\rm dr} + v^{\rm eff} (\p_x e)^{\rm dr} = 0
\eeq
where the dressing operation is defined in \eqref{eq:fdr}. Note that the effective velocity (Eq. 2a in the main text) can be written as
\beq
	v^{\rm eff} = v^{\rm dr} / 1^{\rm dr}
\eeq
where $v^{\rm dr}$ is the dressed of the function $v$ (the identity function), and $1^{\rm dr}$ is the dressed of the constant function 1 ($1^{\rm dr}$ and all dressed quantities are functions of $v$, as well as, of course, the space-time position $x,t$). Therefore, we have
\beq\label{wdr2}
	 (\p_t e)^{\rm dr} + \frc{v^{\rm dr}}{1^{\rm dr}} (\p_x e)^{\rm dr} = 0
\eeq
 
In the initial state, at $t=0$, for each position $x$ the state of the gas is given by the Gibbs ensemble, with a chemical potential $\mu - m\omega_\parallel^2x^2$ and a temperature $T$, and the energy function reduces to~\cite{yangyang}
\beq
	e(x,0,v) = \frc1{T}\lt(\frc{mv^2}2 +\frc{m\omega_\parallel^2 x^2}2 -\mu\rt).
\eeq
We propose that an {\em approximate} solution to the evolution equation from a harmonic potential is of the form
\beq\label{propo}
	e(x,t,v) \approx f(t)v^2 + g(t)vx + h(t) x^2 -\mu.
\eeq
This approximate solution holds if the following condition is approximately satisfied,
\beq\label{cond}
	(v^{\rm dr})^2 \approx (v^2)^{\rm dr} 1^{\rm dr}.
\eeq
Indeed inserting \eqref{propo} in \eqref{wdr} gives
\beq
	f'(t)(v^2)^{\rm dr} + g'(t) v^{\rm dr} x + h'(t) x^2 1^{\rm dr}
	+ g(t) (v^{\rm dr})^2/1^{\rm dr} + h(t) x v^{\rm dr} \approx 0
\eeq
And so we have
\beq
	h'(t) = 0,\quad g'(t) = -h(t),\quad f'(t)(v^2)^{\rm dr} \approx -g(t) (v^{\rm dr})^2/1^{\rm dr}.
\eeq
If the approximation \eqref{cond} holds, then the last (approximate) equation becomes
\beq
	f'(t) \approx -g(t)
\eeq
and we find
\beq\label{sol}
	h(t) = a,\quad g(t) = b-at,\quad
	f(t) = c-bt+at^2/2.
\eeq
The initial condition is satisfied, with $a=m\omega_\parallel^2/(2T)$, $b=0$ and $c =m /(2T)$.

We now note that the approximate solution \eqref{propo} with \eqref{sol} is {\em exactly} the form we would obtain (under \eqref{cond}) from CHD:  in all cells it is a boosted Gibbs state, completely determined by the mass, momentum and energy densities, hence we can restrict to these conservation laws. The validity of CHD thus reduced to the investigation of the validity of  the approximation \eqref{cond}.

From \eqref{w}, we see that $\ep(v)$ has the same leading large-$|v|$ asymptotics as $e(v)$. Hence, in states with power-law growing $e(v)$, the occupation function decays as an exponential of a power at large $|v|$. From the dressing operation \eqref{eq:fdr}, it is then clear that, in any such state, the dressed $f^{\rm dr}(v)$ of any function $f(v)$ has exactly the same asymptotic expansion in powers of $1/v$ up to powers $1/v^2$: the corrections coming from the occupation function are exponentially decaying, hence the main corrections come from integrating the differential scattering kernel around finite regions of the integration variable $w$ (the size of the finite region being controlled by the exponential decay of the occupation function). These give the power law $1/v^2$. That is, assuming $f^{\rm dr}(v)$ stays finite for finite values of $v$, we have
\beq\label{fdrasymp}
	f^{\rm dr}(v) = f(v) + O(1/v^{2}).
\eeq
Therefore
\beq
	(v^{\rm dr})^2 = (v^2)^{\rm dr} 1^{\rm dr} + O(1/v)
\eeq
and the condition \eqref{cond} is satisfied up to vanishing terms at large $v$. This fully justifies that the approximate solution \eqref{propo} indeed holds as written, up to vanishing terms, and shows that CHD reproduces the correct non-vanishing large-$v$ asymptotics of $e(v)$, and thus (by a similar asymptotic analysis as above) of $\ep(v)$.

At small velocities, the effective velocity $v^{\rm eff}$ is different from $v$ due to transport within its local environment: for instance, a quasi-particle of bare velocity $v=0$ acquires a nonzero effective velocity $v^{\rm eff}\neq0$ if its surrounding environment carries nonzero currents. However, these differences are bounded, and tend to be relatively small. Similar effects occur for the dressing of any function, and therefore the approximation \eqref{cond} has an error that stays bounded at finite $v$.

Finally, we note from the above calculations and especially from \eqref{fdrasymp} that the {\em free evolution} (with trivial dressing) generically (that is, in all situations where the occupation function decays at least exponentially) describes correctly the first three leading powers of the large-$v$ asymptotics of the source term $e(v)$ under the one-dimensional expansion (recall that in the case of the initial harmonic potential, the free evolution agrees with CHD, but that in general it does not). One could for instance consider an initial potential with higher powers, say up $x^4$. Then we see that an approximate solution of similar type, valid at large $v$, would have powers of $v$ up to $v^4$, with the coefficients of the powers $v^2$, $v^3$ and $v^4$ correctly described by a free evolution equation. This is definitely very far from a CHD solution, as powers of $v^3$ and $v^4$ are forbidden in CHD. That is, when starting from non-harmonic potentials, CHD does not describe the correct large-$v$ asymptotics of the distribution, hence is quite far for all $v$ -- the maximal deviation for the function $e(v)$ is in fact is unbounded. It is these effects that are seen in section \ref{sectII}.


\section{GHD vs. CHD: comparison of the phase-space occupation functions $\nu(x,v)$}\label{sectII}

Here we display the phase-space occupation $\nu (x,v)$ for the expansion from a double-well potential shown in Fig. 3 in the main text. At each $x$, the occupation function $\nu(v)$ is related to the density $\rho(v)$ used in the main text by Eq. (\ref{eq:nu_rho}). At thermal equilibrium, one can think of it as a Fermi distribution: at zero temperature, it is exactly a rectangular function, with support $[-2\sqrt{\frac{g n}{m}},2\sqrt{\frac{g n}{m}}]$ in the limit of weak repulsion. At finite temperature the two sides of the rectangle are rounded off. The GHD evolution equation can be written directly in terms of the occupation function $\nu (v)$, see Eq.~(\ref{eq:GHD_nu}).

On the phase-space pictures in Fig.~\ref{figsupp:ghd_vs_chd}, one clearly sees why CHD must differ from GHD. Indeed, initially the phase-space occupation is the equilibrium one at temperature $T=0.3\, \mu{\rm K}$, obtained from LDA in the double-well potential $V(x)$, see the main text. Because the atomic density increases near the two minima of the potential $V(x)$, the interval $[v_1,v_2]$ in which the occupation function is close to $1$ is larger near these points. This results in the support of $\nu(x,t)$ looking like a peanut at $t=0$. By definition, the initial distribution is the same for GHD and CHD.
\begin{center}
\begin{figure}[h]
\begin{tabular}{lll}
Generalized HydroDynamics (GHD): \\
	\includegraphics[width=0.33\textwidth]{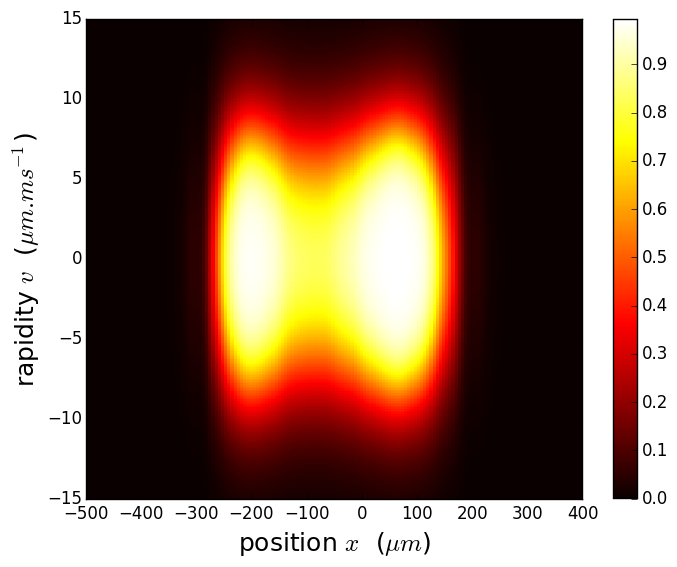} &
	\includegraphics[width=0.33\textwidth]{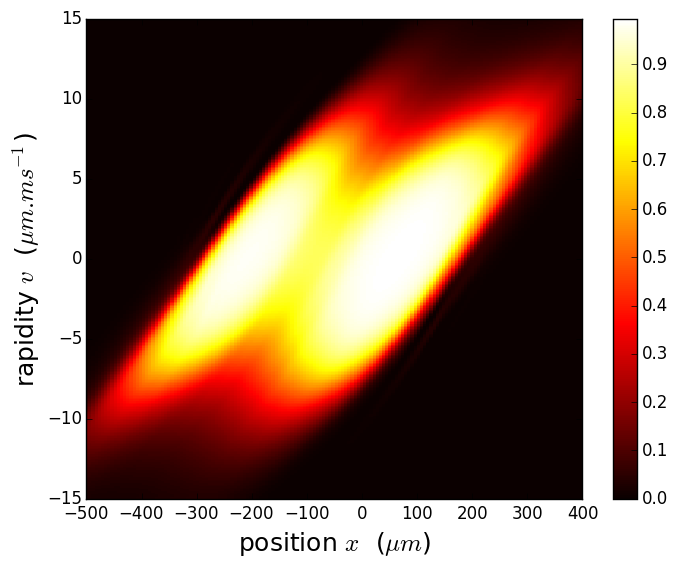} &
	\includegraphics[width=0.33\textwidth]{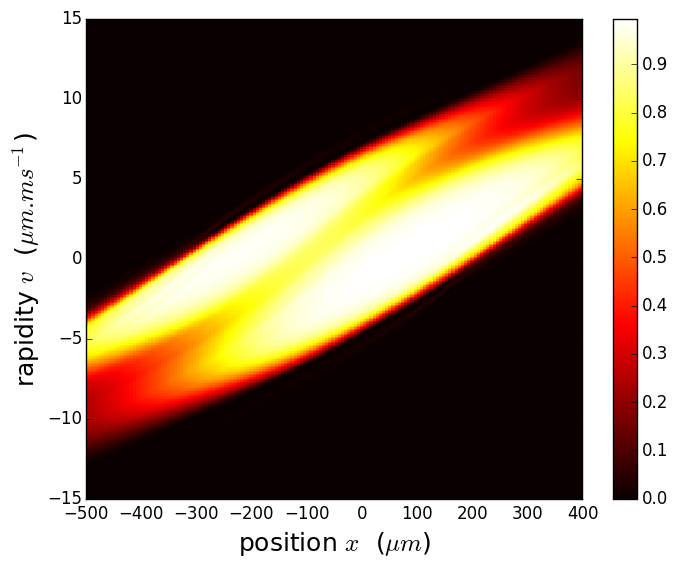} \\
    \qquad $t=0$ & \qquad $ t= 25$ ms & \qquad $t = 55$ ms \\
Conventional hydrodynamics (CHD): \\
	\includegraphics[width=0.33\textwidth]{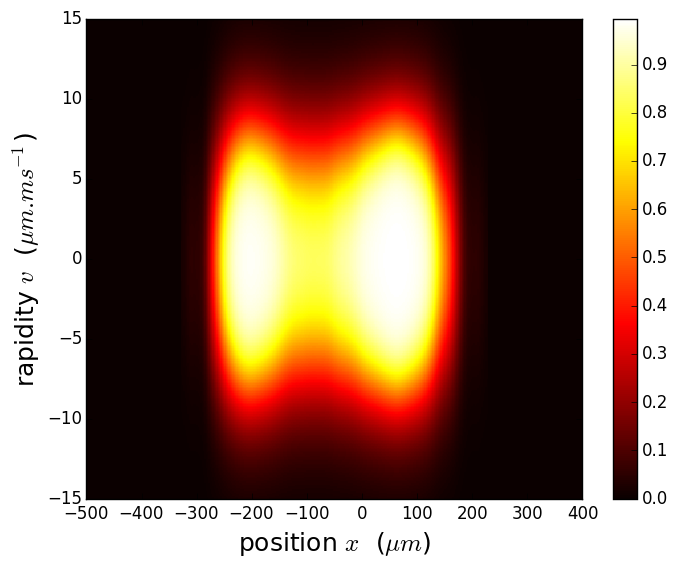} &
	\includegraphics[width=0.33\textwidth]{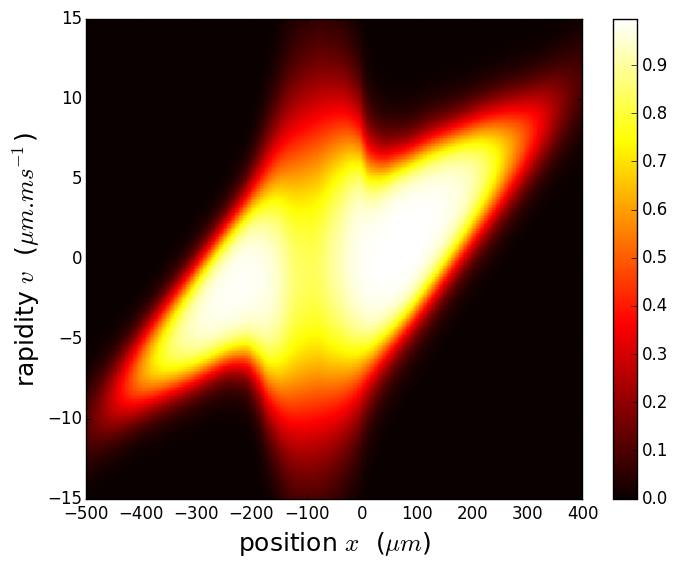} &
	\includegraphics[width=0.33\textwidth]{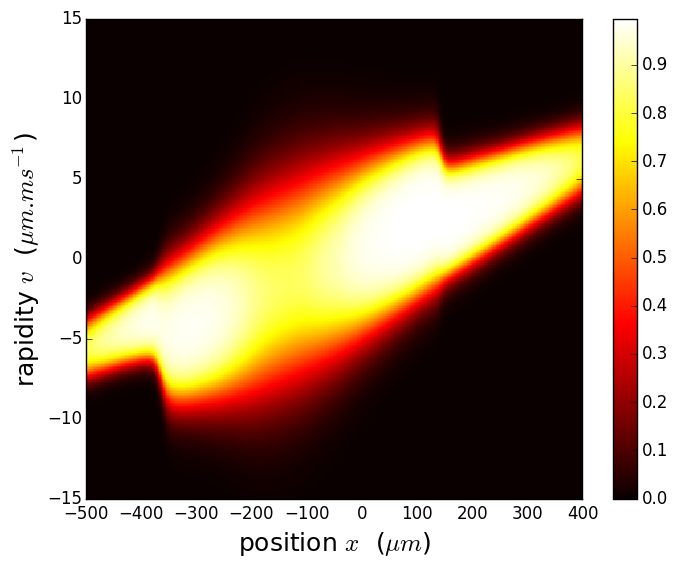} \\
\qquad $t=0$ & \qquad $ t= 25$ ms & \qquad $t = 55$ ms
\end{tabular}
	\caption{Phase-space occupation $\nu(x,v)$ at time $t$ for the parameters of Fig.~3 in the main text, simulated with GHD and CHD. Since CHD does not allow non-thermal local distributions of rapidities, we see that the distribution gets quickly distorted, compared to the GHD one. This results in the two in situ density profiles being clearly different, see Fig.~3 in the main text.}
\label{figsupp:ghd_vs_chd}
\end{figure}
\end{center}
At later times, the distribution gets distorted. GHD predicts that the top of the support of $\nu(x,v)$ moves to the right, while the bottom of the support moves to the left. At $t \simeq 25$ ms we already see that, for a fixed $x$ in the central region around $x=0$, the occupation function $\nu(v)$ does not look like a thermal equilibrium distribution anymore: instead of being a rounded rectangular function, it is now a function that has two maxima. This gets worse as time increases: the distribution $\nu(v)$, for fixed position $x$, differs more and more from an equilibrium one.

CHD, on the contrary, enforces an equilibrium occupation function at all positions and all times. Therefore, it must differ significantly from GHD. Indeed, this is what we see in the phase-space occupation at $t=25$ ms, where the occupation function of CHD is distorted in the central region, compared to the GHD one, in order to maintain a thermal state. The discrepancy between GHD and CHD then keeps increases at later times.

The density profiles of Fig. 3 in the main text are obtained from these phase-space occupation function $\nu (x, \theta)$ by Eq. (\ref{eq:nu_rho}) and then integrating over all rapidities at fixed position $x$, $n(x) = \int \rho(x,v) dv$.

\newpage

\begin{figure}[htb]
\centerline{\includegraphics{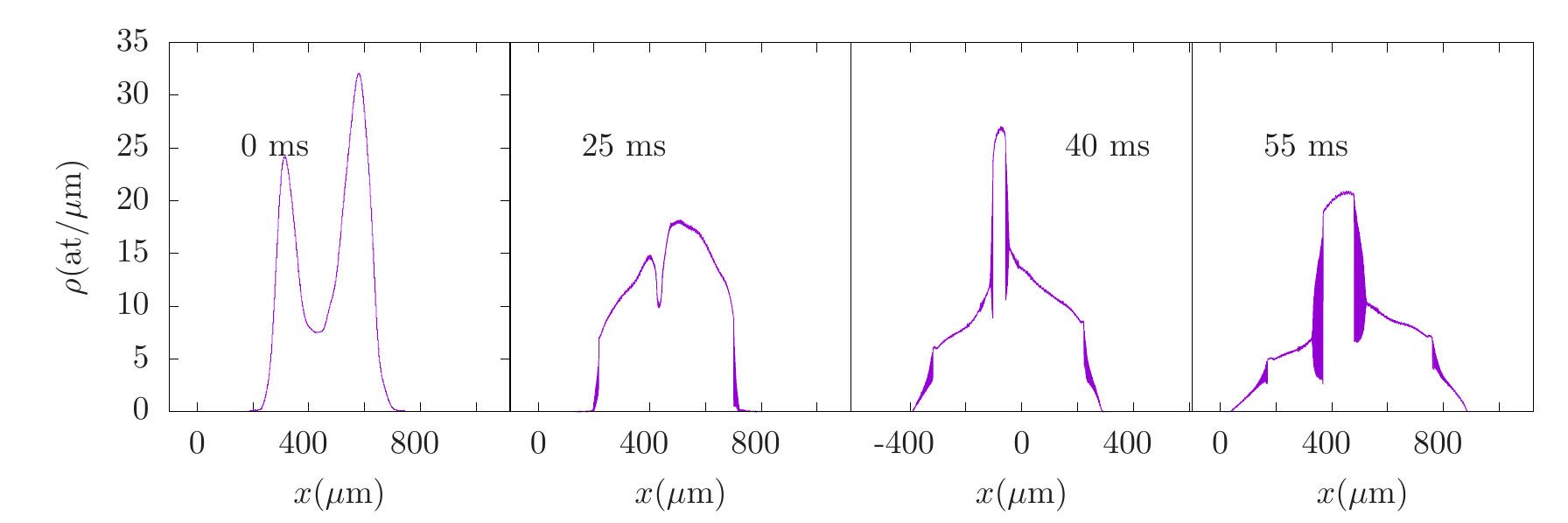}}
\caption{Expansion of a cloud whose initial density distribution is 
equal to that of the experimental data of Fig.~3 in the main text,
according to the Gross-Pitaevskii equation.
The evolution is very different from that observed experimentally, with the appearance 
of a central peak, not seen experimentally.}
\label{fig.GPexpansion}
\end{figure}

\section{Gross Pitaevskii predictions for expansion from a double well}
We performed a Gross Pitaevskii calculation  for the situation considered in Fig.3. 
In this calculation, the initial wavefunction is $\psi(x)=\sqrt{n_0(x)}$, where
$n_0(x)$ is the initial experimental profile. 
We then evolve this initial profile according to the 
time-dependant Gross-Pitaevskii equation Eq.~(\ref{eq.GP}). 
The resulting time evolution, shown in Fig.(\ref{fig.GPexpansion}), is very different from that 
observed experimentally. This indicates that thermal excitations initially present in the cloud play an important role in the time-evolution shown in Fig.3. Note that GHD calculations performed at a very low temperature are in agreement with these Gross-Pitaevskii calculations, provided fast oscillations shown in the Gross-Pitaevski profiles are averages out (see section \ref{sec.runningGHDwithoutV}).

\newpage


\section{Running the GHD simulation without good knowledge \\ of the initial potential $V(x)$}
\label{sec.runningGHDwithoutV}

\begin{center}
\begin{figure}[h]
	\begin{tabular}{cc}
	\includegraphics[width=0.45\textwidth]{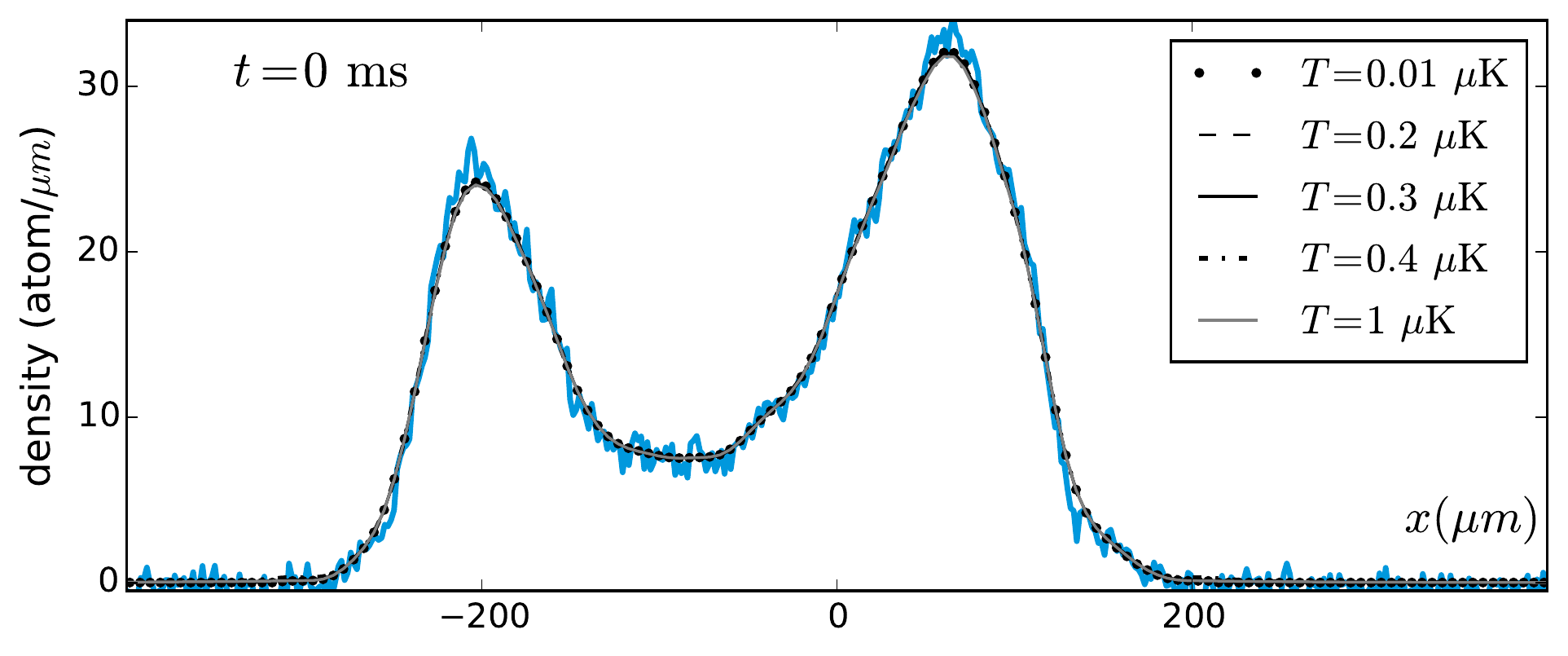} &
	\includegraphics[width=0.45\textwidth]{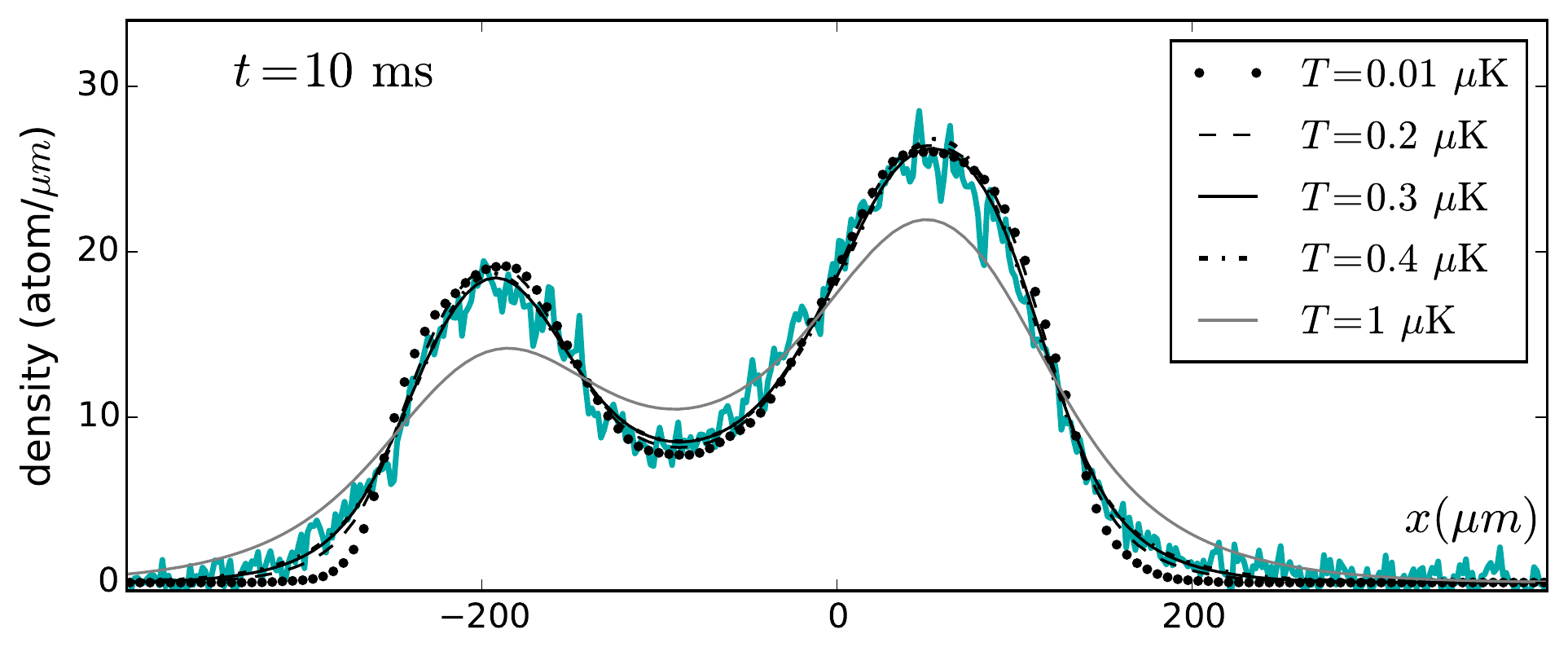} \\
	\includegraphics[width=0.45\textwidth]{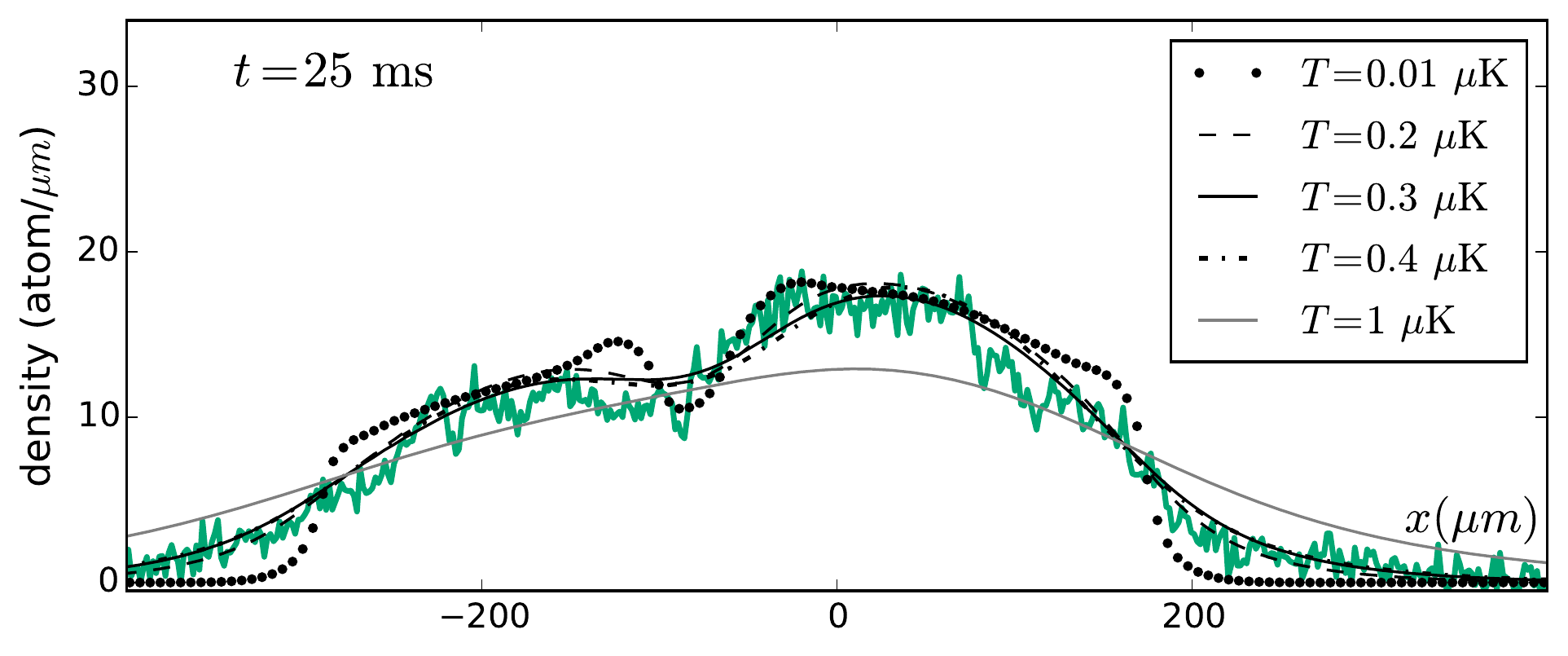} &
	\includegraphics[width=0.45\textwidth]{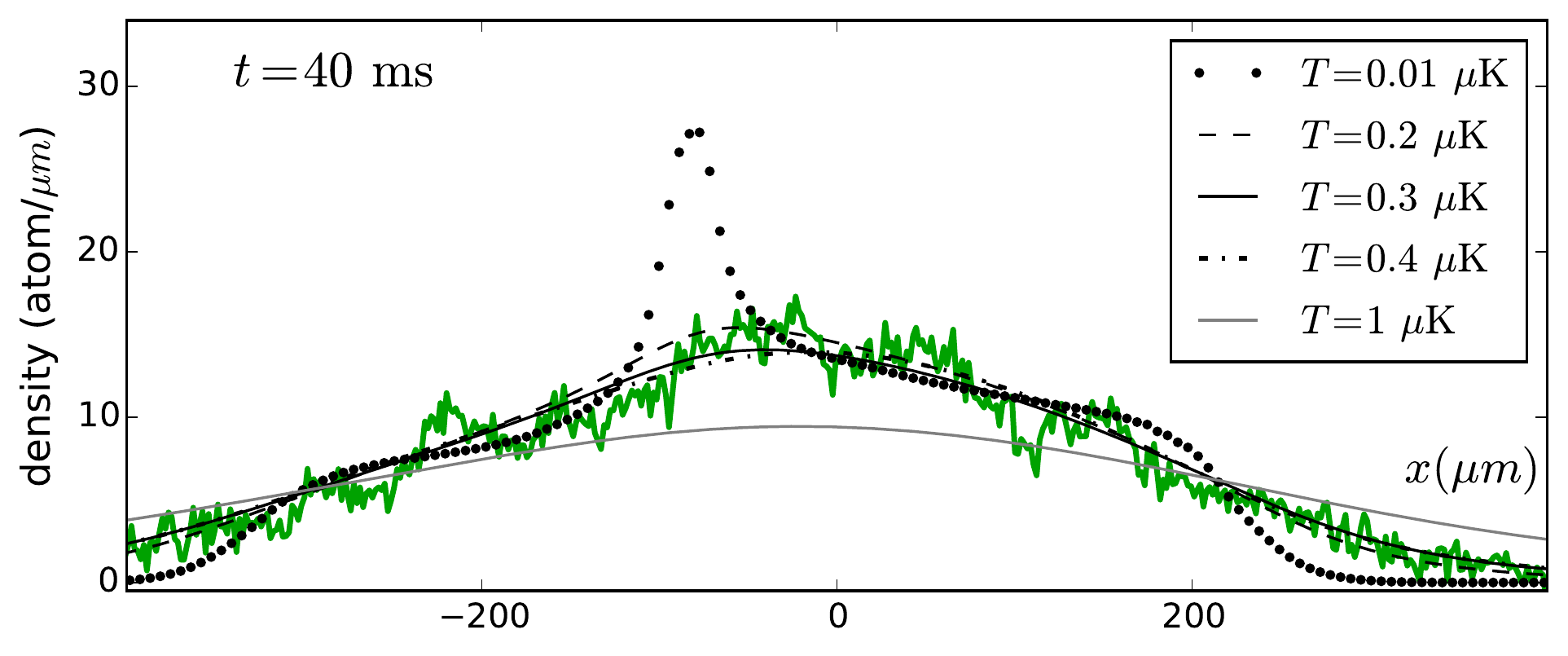} \\
	\includegraphics[width=0.45\textwidth]{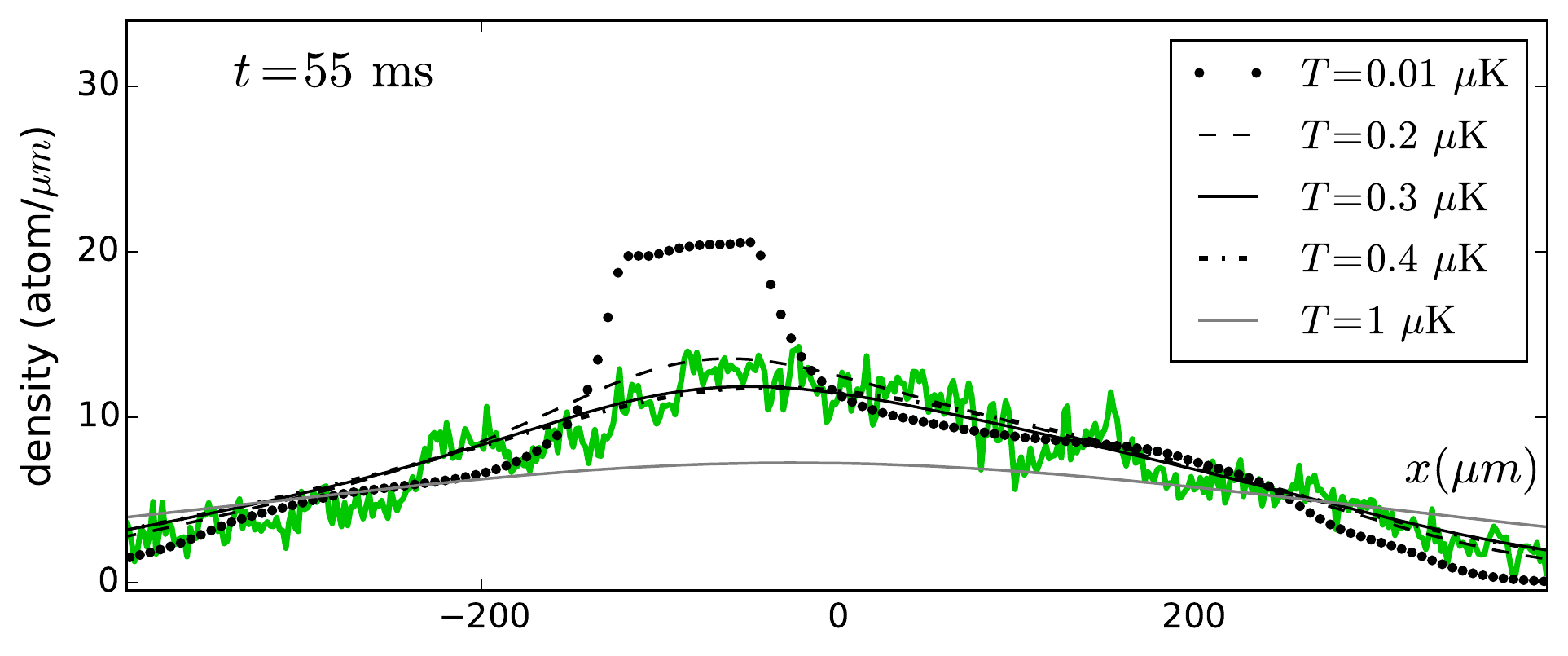}
	\end{tabular}
    \caption{Comparison between the experimental data of Fig.~3 in the main text, and the GHD simulation assuming three different temperatures of the initial state. For Fig.~3 in the main text, we selected the results corresponding to $T = 0.3 \,\mu$K.}
    \label{figsupp:diffTfig3}
\end{figure}
\end{center}

The creation of a double-well potential on the chip is experimentally tricky, and, as a result, we do not have a very good knowledge
of this potential. It should be close to a polynomial potential of fourth degree, $V(x) \simeq \sum_{p=0}^4 a_p x^p$, but we do not know the coefficients $a_p$. One could extract them from a fit, but then we would have $6$ adjustable parameters in total: the five coefficients $a_p$, and the temperature of the initial state.

On the other hand, the initial potential $V(x)$ is needed only to reconstruct the initial distribution $\rho(x,v)$. So, instead of trying to reconstruct first the potential $V(x)$ and then the initial distribution $\rho(x,v)$, we find it more satisfying to construct directly that distribution, assuming that it is given by a thermal Gibbs state in every fluid cell at position $x$. When we do that, there is a single adjustable parameter: the initial temperature, assumed to be initially homogeneous throughout the cloud.

We proceed as follows. First, we take a convolution of the experimental density at time $t=0$ with a gaussian, in order to get a smooth initial density profile $n(x)$. Then we fix arbitrarily a temperature $T$, and, assuming that the gas is locally at equilibrium at temperature $T$, we reconstruct the occupation function $\nu (x, \theta)$---or equivalently the distribution $\rho(x,\theta)$---using the Yang-Yang equation of state (\ref{eq:TBA}). Let us emphasize that, at this point, we have reconstructed a density profile, which  corresponds to some potential $V_T(x)$, and the latter potential is not independent of the choice of the temperature $T$. What we then want, of course, is to identify the correct $T$ corresponding to the correct experimental $V(x)$ which, unfortunately, we do not know a priori. We then run the GHD simulation, which allows us to compute the density profiles at later times $t$. Those density profiles at later times depend on the initial state, therefore they depend on $T$.

\begin{center}
\begin{figure}[h]
	\begin{tabular}{cc}
	\includegraphics[width=0.45\textwidth]{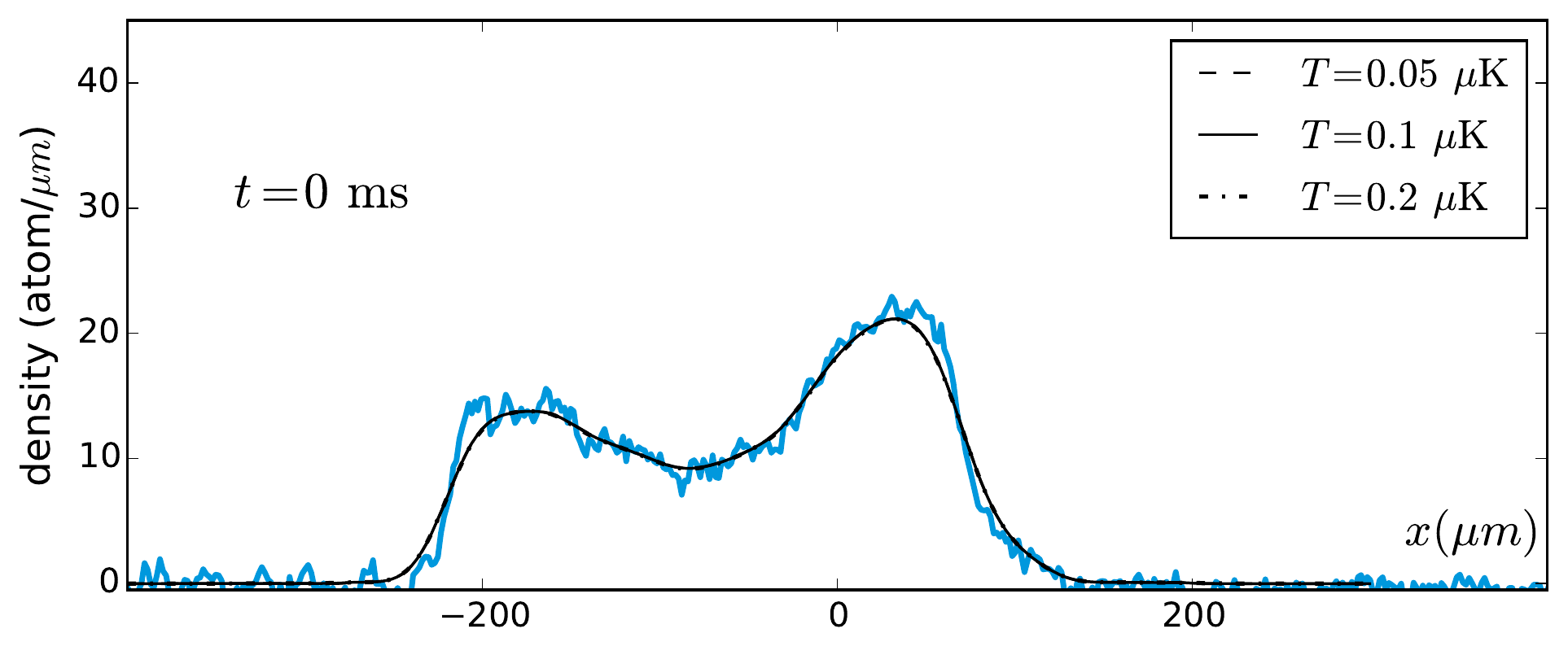} &
	\includegraphics[width=0.45\textwidth]{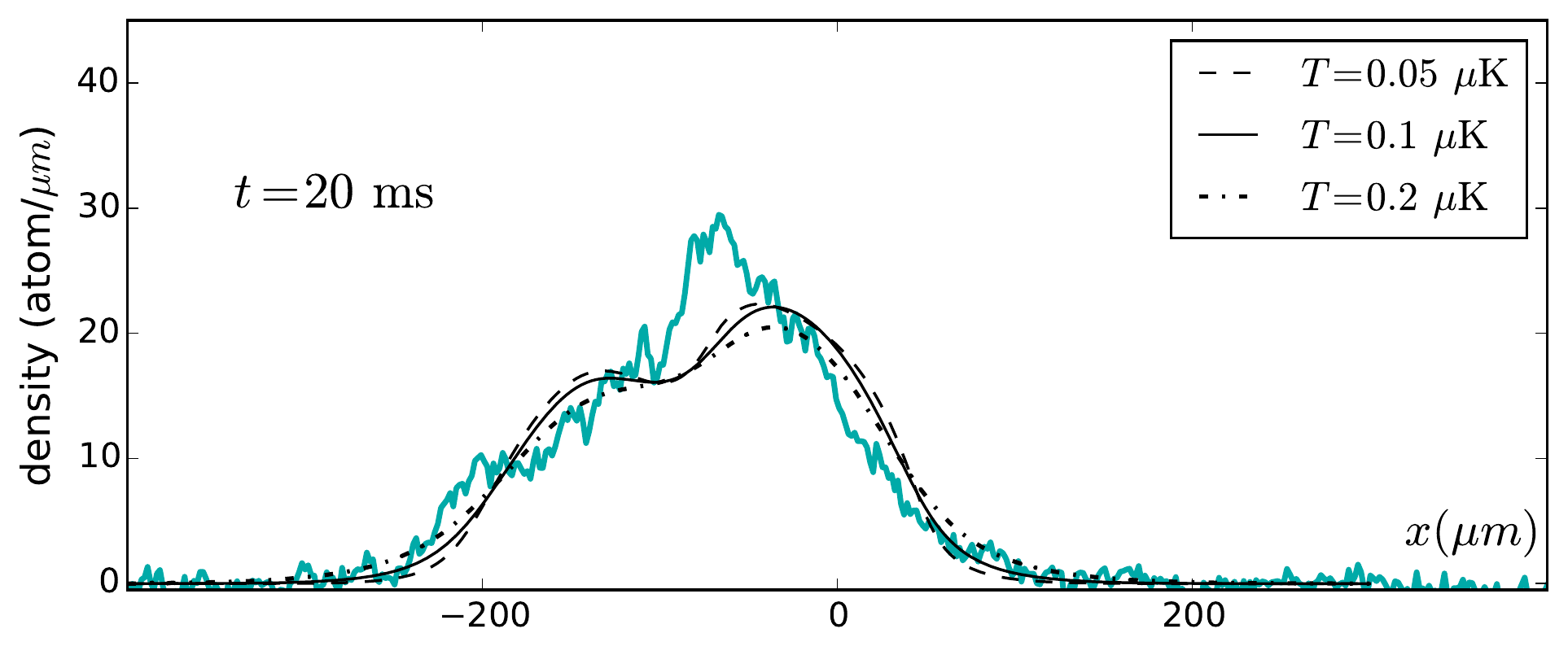} \\
	\includegraphics[width=0.45\textwidth]{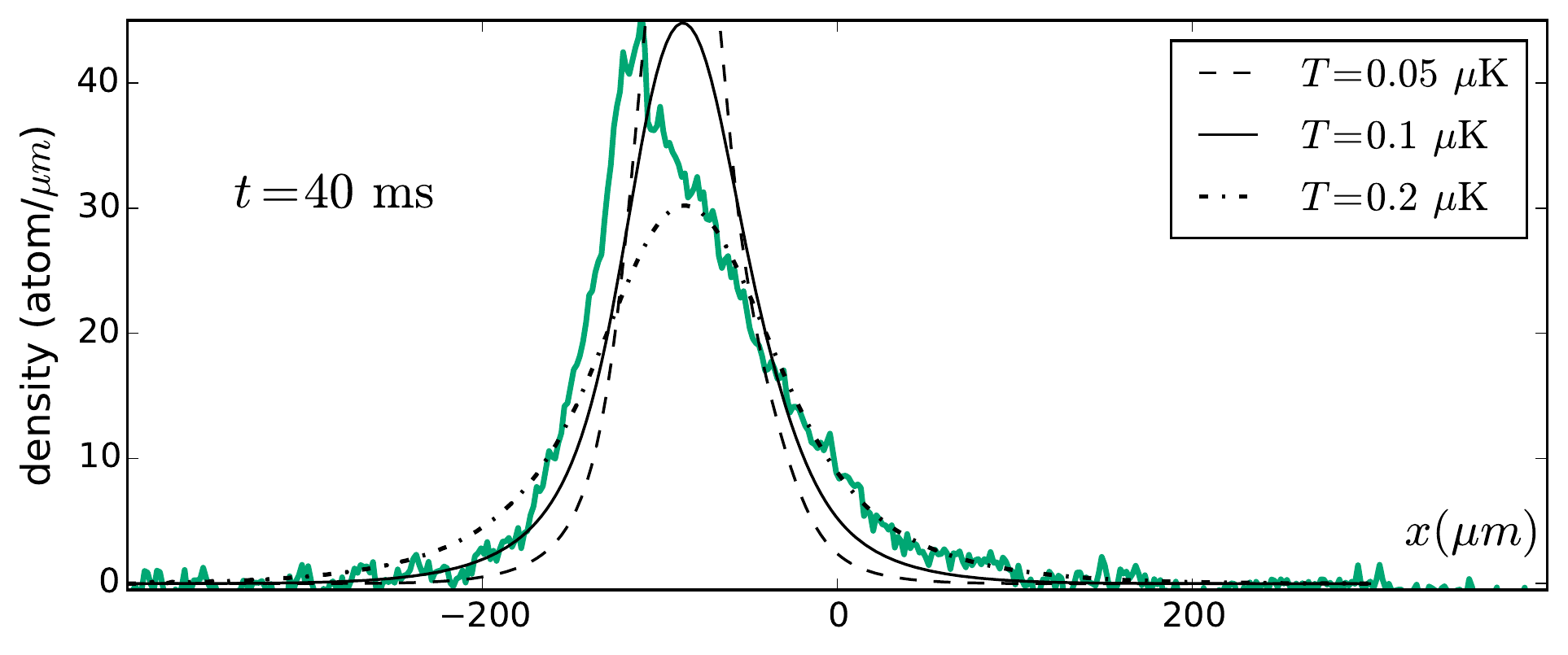} &
	\includegraphics[width=0.45\textwidth]{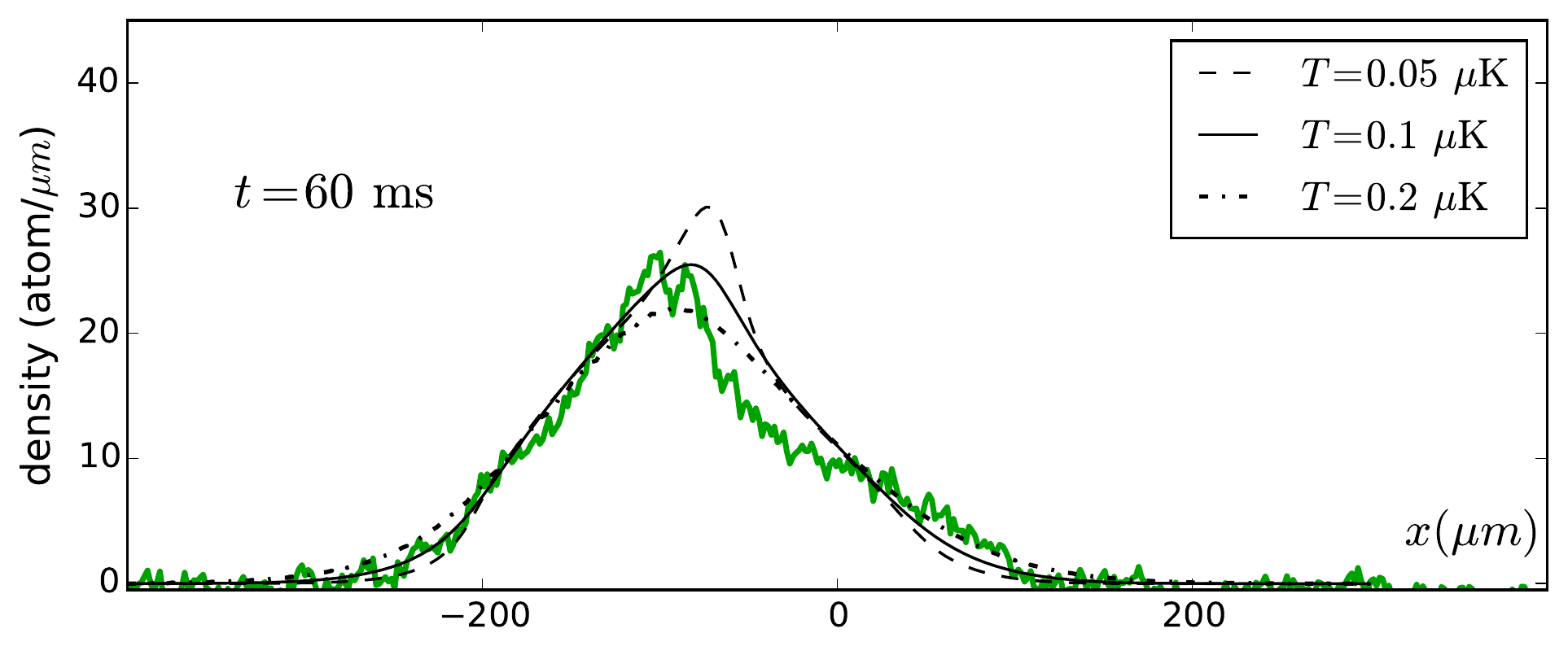} \\
	\includegraphics[width=0.45\textwidth]{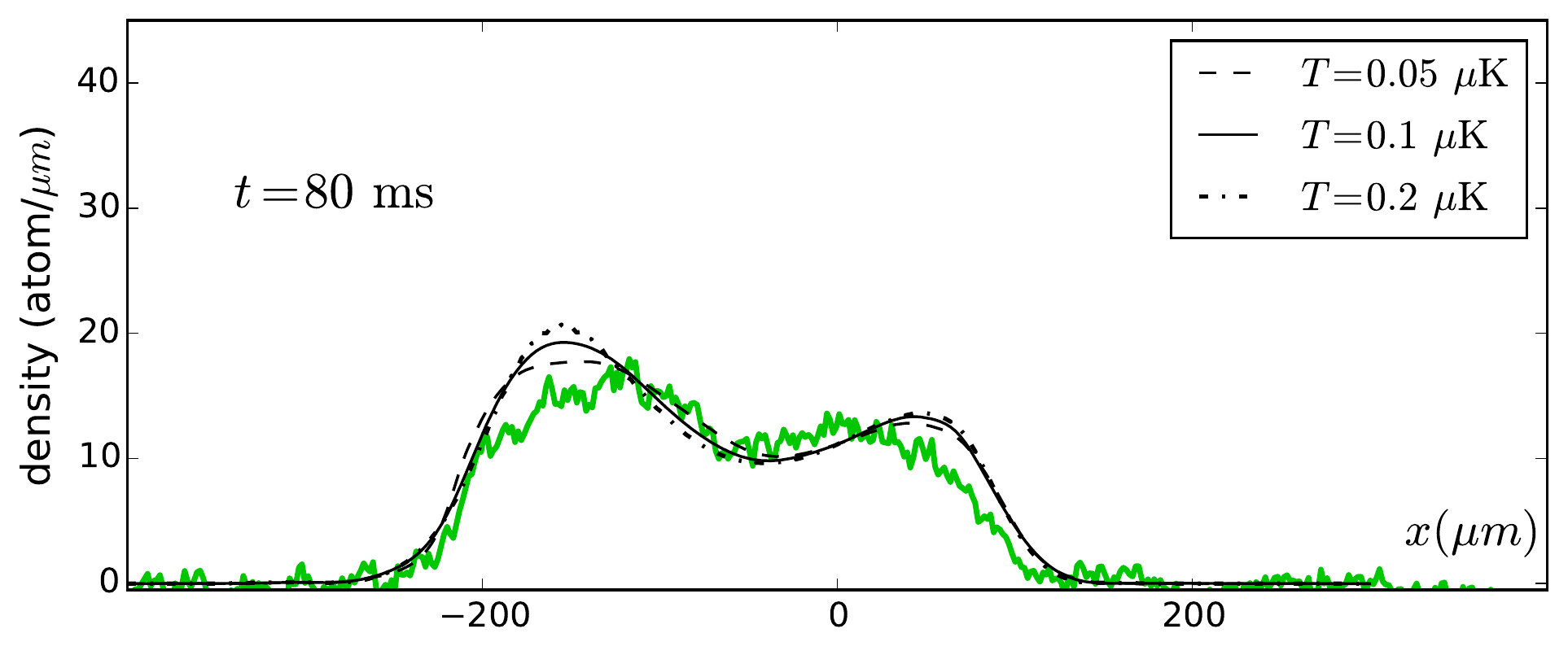} &
	\includegraphics[width=0.45\textwidth]{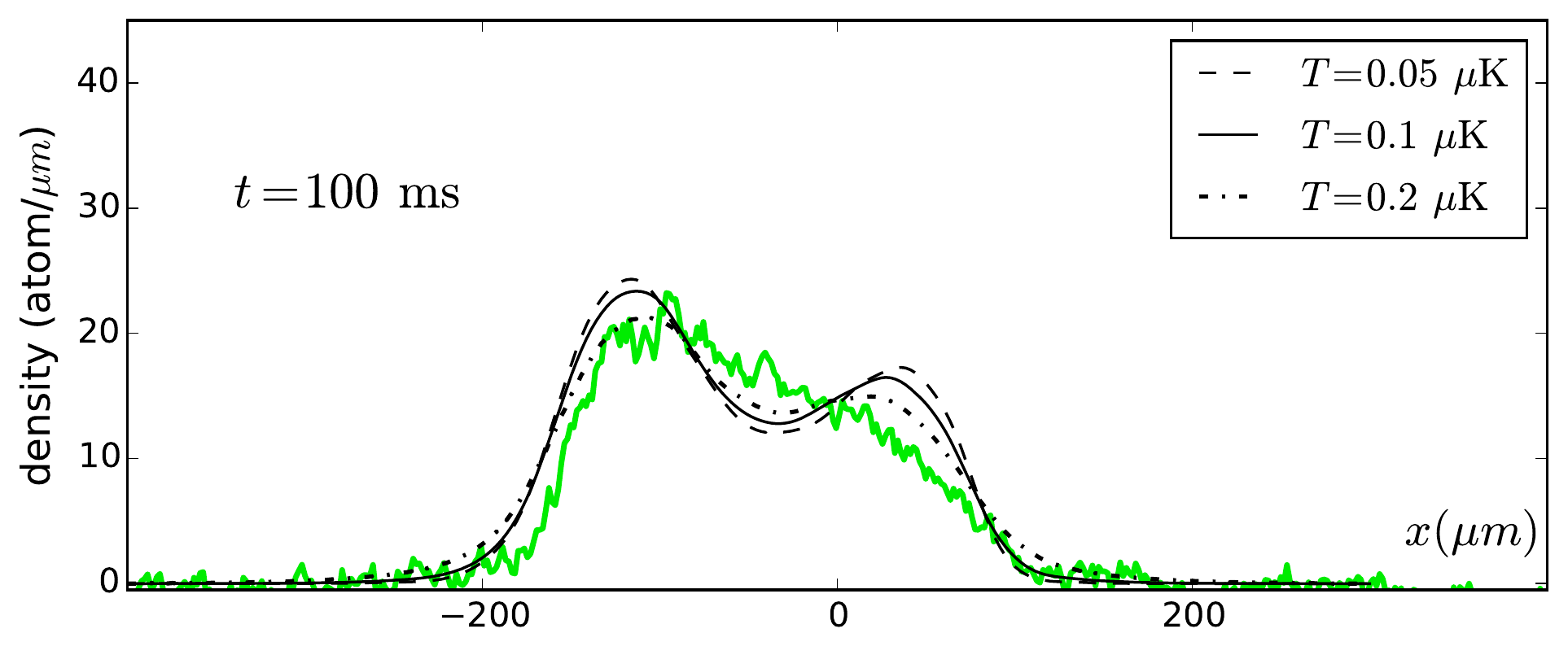}
	\end{tabular}
    \caption{Comparison between the experimental data of Fig.~4 in the main text, and the GHD simulation assuming three different temperatures of the initial state. For Fig.~4 in the main text, we selected the results corresponding to $T = 0.1 \,\mu$K.}
    \label{figsupp:diffTfig4}
\end{figure}
\end{center}

We do this for several values of the initial temperature $T$. In Fig.~\ref{figsupp:diffTfig3} we display the results for the data set shown in Fig. 3 in the main text, for initial temperatures $0.2 \,\mu{\rm K}$, $0.3 \,\mu{\rm K}$ and $0.4 \,\mu{\rm K}$. We see that the results do not strongly dependent on the temperature. More quantitatively, for each $T$ we measure the mean distance 
between the experimental data and the GHD prediction defined as $\sqrt{\frac{1}{N_{\rm{data}}}\sum_{x,t} [n_{\rm data}(x,t)-n_{\rm GHD}(x,t)]^2}$, where the sum 
is done over all the data points 
for different positions $x$ and times $t$ and $N_{\rm{data}}$ is the total 
number of data points. For temperatures $T = 0.01, 0.2, 0.25, 0.3, 0.35, 0.4, 1  \,\mu{\rm K}$, we find that the distance is respectively $1.21, 0.66, 0.62, 0.61, 0.64, 0.63, 1.26$ (atom/$\mu$m). In the main text, we have selected the GHD results with $T = 0.3 \,\mu{\rm K}$, corresponding to the minimum of that distance for the set of temperatures simulated.

We have proceeded in the same way for the quench from a double-well to harmonic potential, corresponding to the data set shown in Fig. 4 in the main text. In Fig.~\ref{figsupp:diffTfig4} we display the GHD curves for initial temperatures $0.05 \,\mu{\rm K}$, $0.1 \,\mu{\rm K}$ and $0.2 \,\mu{\rm K}$, up to time $t = 100$ ms after the quench. We have performed the GHD simulation for temperatures For temperatures $T = 0.05, 0.75, 0.1, 0.125, 0.15,0.2,0.3  \,\mu{\rm K}$, we find that the mean distance  is respectively $2.69, 2.19, 1.88, 1.71, 1.64, 1.72, 2.08$ (atom/$\mu$m). In the main text, we have selected the GHD results with $T = 0.15 \,\mu{\rm K}$, corresponding to the minimum of that distance for the set of temperatures simulated.

\section{Approximate periodicity of the GHD solution for the quench from double-well to harmonic potential}

In Fig.~\ref{figsupp:period} we display the phase-space occupation $\nu (x,v)$ for the quench from double-well to harmonic potential, corresponding to the GHD data shown in Fig. 4 in the main text. We recall that $\nu(x,v)$ is the phase-space occupation, and that it is related to the phase-space density used in the main text by Eq.~(\ref{eq:nu_rho}). One can roughly think of $\nu(x,v)$ as a Fermi dstribution, see App. III above. The GHD evolution equation can be written directly in terms of the occupation function $\nu (v)$, see Eq.~(\ref{eq:GHD_nu}).

It is clear from these plots that, after one period of the harmonic trap (the period is 154~ms), the distribution is only approximately similar to the initial one. So, even though our atomic clouds are weakly interacting, the behavior that is predicted by GHD is clearly different from the one of non-interacting particles, since the latter would be exactly periodic.
\begin{center}
\begin{figure}[h]
\begin{tabular}{llll}
	\includegraphics[width=0.25\textwidth]{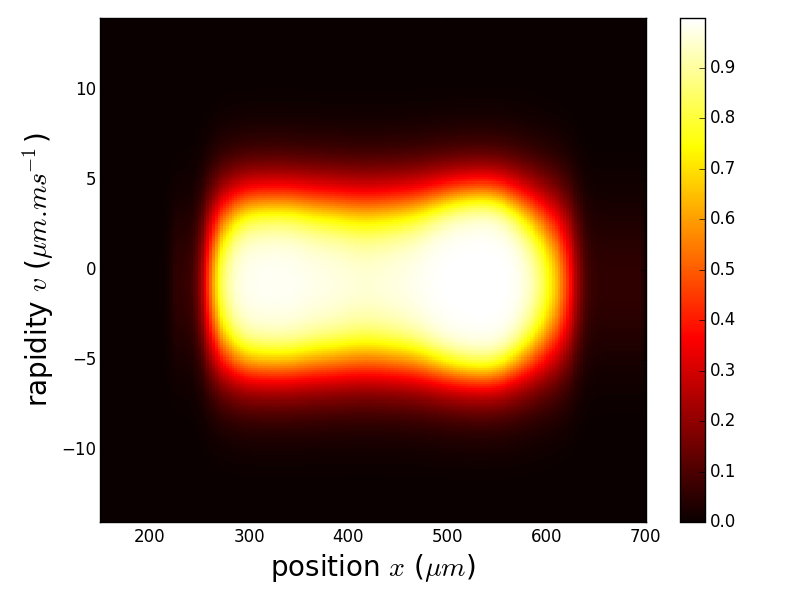} &
	\includegraphics[width=0.25\textwidth]{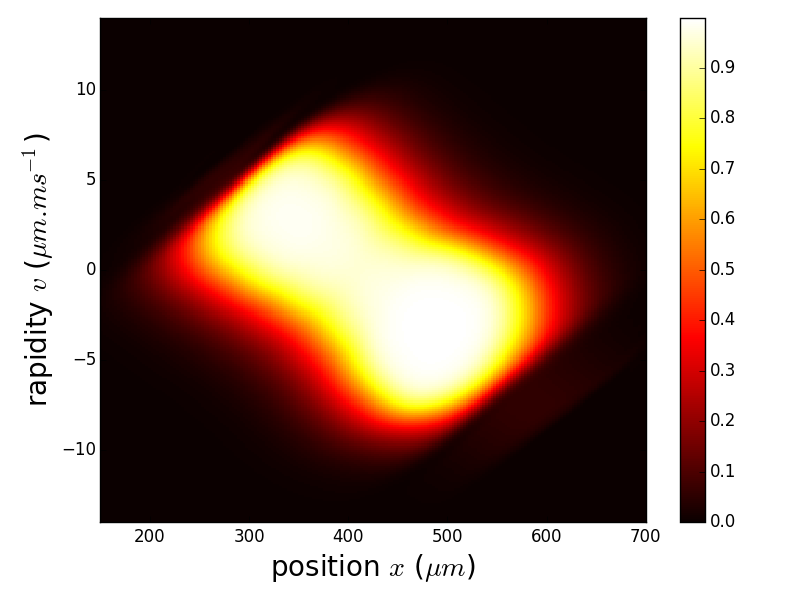} &
	\includegraphics[width=0.25\textwidth]{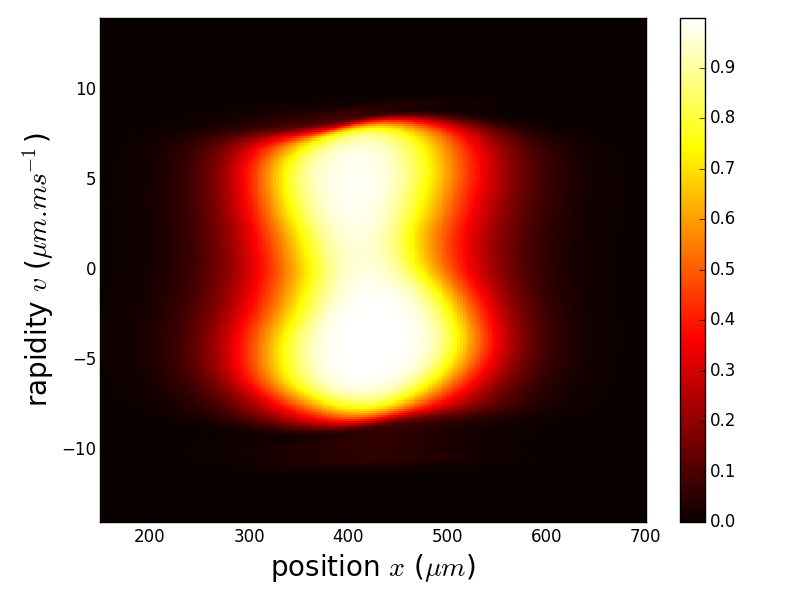} &
	\includegraphics[width=0.25\textwidth]{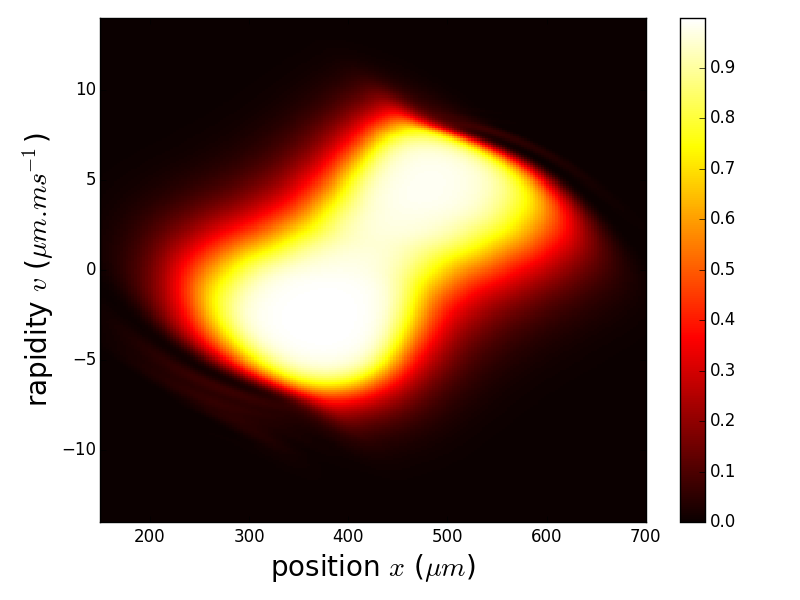} \\
    \qquad $t=0$ & \qquad $ t= 20$ ms & \qquad $t = 40$ ms & \qquad $t = 60$ ms \\ \\
	\includegraphics[width=0.25\textwidth]{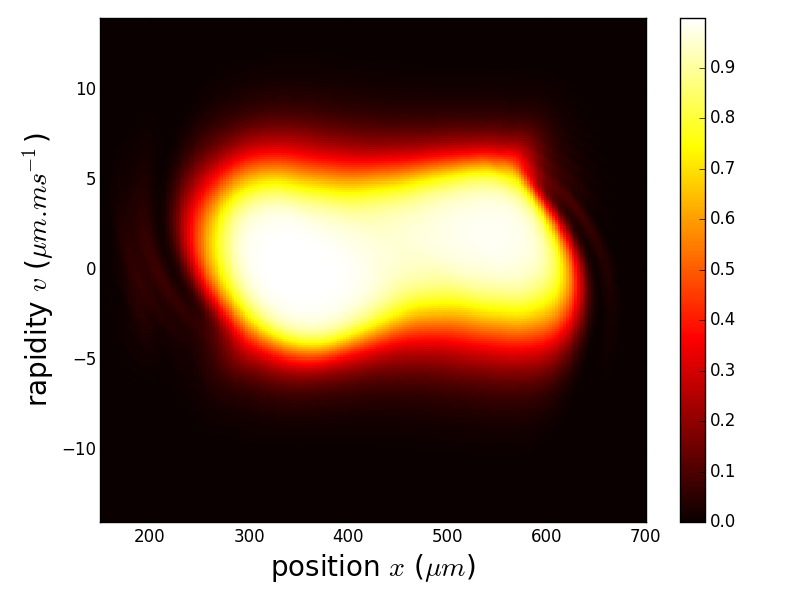} &
	\includegraphics[width=0.25\textwidth]{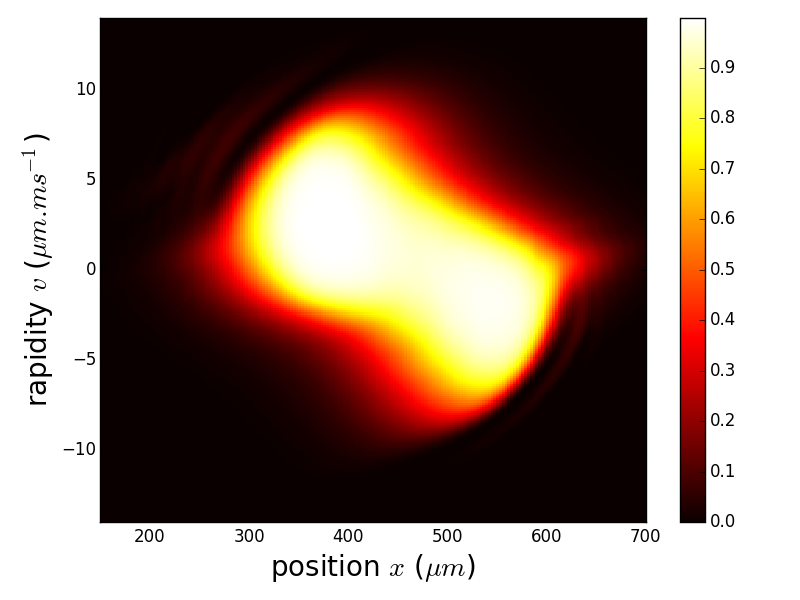} &
	\includegraphics[width=0.25\textwidth]{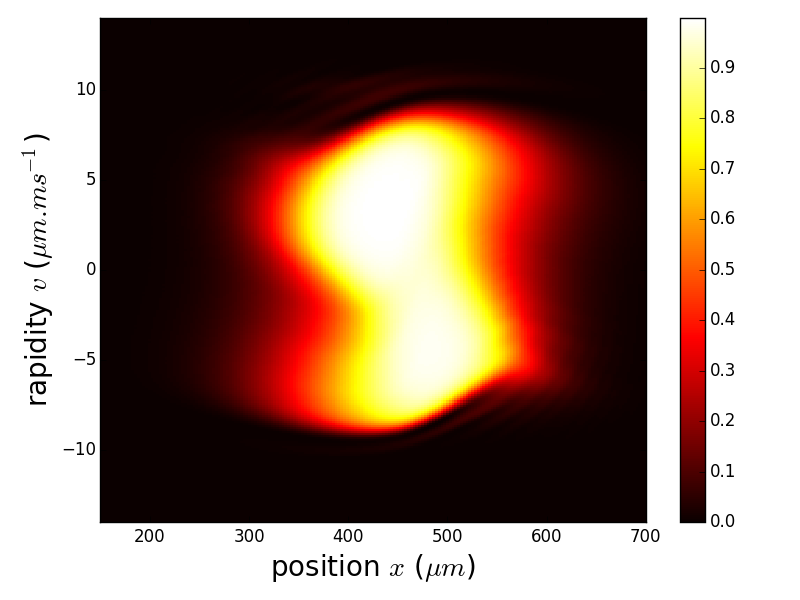} &
	\includegraphics[width=0.25\textwidth]{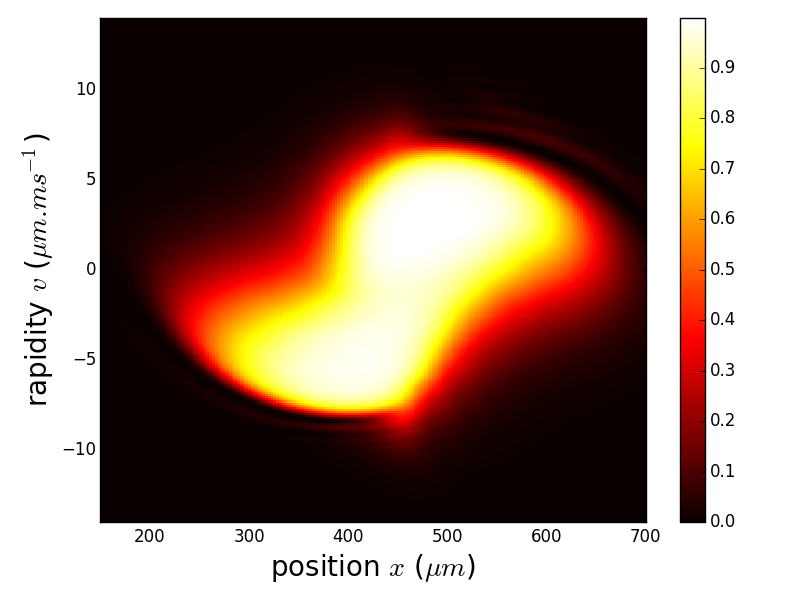} \\
    \qquad $t=80$ ms & \qquad $ t= 100$ ms & \qquad $t = 120$ ms & \qquad $t = 140$ ms \\ \\
	\includegraphics[width=0.25\textwidth]{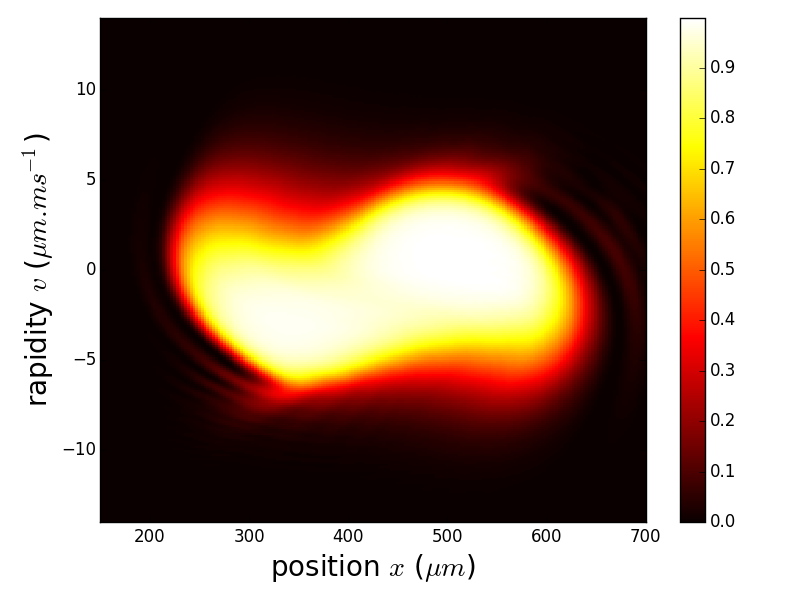} &
	\includegraphics[width=0.25\textwidth]{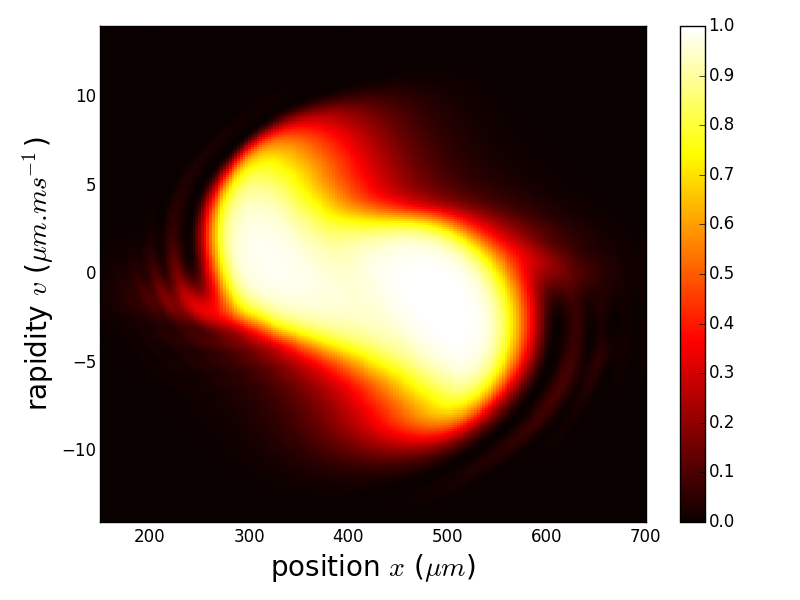} & \\
    \qquad $t=160$ ms & \qquad $ t= 180$ ms  
\end{tabular}
	\caption{Phase-space occupation $\nu(x,v)$ obtained from GHD for the quench from double-well to harmonic potential shown in Fig.~4 in the main text. The motion is {\it approximately} periodic, although one clearly sees that the distribution does not come back exactly to its initial shape after one period; instead, it gets slightly distorted by the interactions between the atoms. The period of the harmonic trap is 154\,ms, and we see that the approximate period is roughly 165~ms.}
    \label{figsupp:period}
\end{figure}
\end{center}

\end{widetext}


\begin{thebibliography}{99}
\bibitem{hanssen} J.-P. Hansen and I.R. McDonald, ``Theory of simple liquids'', Second Edition. Academic Press, San Diego CA, 1990.

\bibitem{allen} M.P. Allen and D.J. Tildesley, ``Computer simulation of liquids'', Second Edition. Oxford University Press, Oxford UK, 2017.

\bibitem{spohn_book} H. Spohn, ``Large Scale Dynamics of Interacting Particles'', Springer, Berlin, 1991.

\bibitem{percus} J.K. Percus, ``Exact solution of kinetics of a model classical fluid'', The Physics of Fluids, 12(8), 1560-1563 (1969).

\bibitem{dobrods} C. Boldrighini, R. L. Dobrushin and Yu. M. Sukhov, ``One-dimensional hard rod caricature of hydrodynamics'', J. Stat. Phys. {\bf 31}, 577 (1983).

\bibitem{dyc} B. Doyon, T. Yoshimura and J.-S. Caux, ``Soliton gases and generalized hydrodynamics'', Phys. Rev. Lett. {\bf 120}, 045301 (2018).

\bibitem{ghd} O. A. Castro-Alvaredo, B. Doyon, and T. Yoshimura, ``Emergent Hydrodynamics in Integrable Quantum Systems Out of Equilibrium'', Phys. Rev. X {\bf 6}, 041065 (2016).
\bibitem{bertini1} B. Bertini, M. Collura, J. De Nardis, and M. Fagotti, ``Transport in Out-of-Equilibrium XXZ Chains: Exact Profiles of Charges and Currents'', Phys. Rev. Lett. {\bf 117}, 207201 (2016).

\bibitem{Amerongen-Yang-2008} A.H. van Amerongen, J.J.P. van Es, P. Wicke, K.V. Kheruntsyan, and N.J. van Druten, ``Yang-Yang Thermodynamics on an Atom Chip'', Phys. Rev. Lett. {\bf 100}, 090402 (2008).

\bibitem{Jacqmin-Sub-Poissonian-2011} 
T. Jacqmin, J. Armijo, T. Berrada, K.V. Kheruntsyan, I. Bouchoule, ``Sub-poissonian fluctuations in a 1d bose gas: From the quantum quasicondensate to the strongly interacting regime'', Phys. Rev. Lett. {\bf 106}, 230405 (2011).

\bibitem{Volger-Thermodynamics-2013} A. Vogler, R. Labouvie, F. Stubenrauch, G. Barontini, V. Guarrera, and H. Ott, ``Thermodynamics of strongly correlated one-dimensional Bose gases'', Phys. Rev. A {\bf 88}, 031603(R) (2013).

\bibitem{LiebLiniger} E.H. Lieb, W. Liniger, ``Exact analysis of an interacting Bose   gas. I. The general solution and the ground state'', Phys. Rev. {\bf 130}(4), 1605 (1963); F. Berezin, G. Pokhil, V. Finkelberg, ``Schr\"odinger equation for a system of 1d particles with point interaction'', Vestnik MGU {\bf 1}, 21 (1964).

\bibitem{dy} B. Doyon, T. Yoshimura, ``A note on generalized hydrodynamics: inhomogeneous fields and other concepts", SciPost Phys. {\bf 2}, 014 (2017).

\bibitem{ddky} B. Doyon, J. Dubail, R. Konik and T. Yoshimura, ``Large-scale description of interacting one-dimensional Bose gases: generalized hydrodynamics supersedes conventional hydrodynamics'', Phys. Rev. Lett. 119, 195301 (2017).

\bibitem{doyon_spohn} B. Doyon and H. Spohn, ``Drude weight for the Lieb-Liniger Bose gas'', SciPost Physics, 3(6), 039 (2017).

\bibitem{Berkeley_prl}  V.B. Bulchandani, R. Vasseur, C. Karrasch, J.E. Moore, ``Solvable Hydrodynamics of Quantum Integrable Systems'', Phys. Rev. Lett. 119, 220604 (2017).

\bibitem{ghd_qnc} J.-S. Caux, B. Doyon, J. Dubail, R. Konik and T. Yoshimura, ``Hydrodynamics of the interacting Bose gas in the Quantum Newton Cradle setup'', arXiv:1711.0873.

\bibitem{viewpoint} J. Dubail, ``A More Efficient Way to Describe Interacting Quantum Particles in 1D'', Viewpoint on Refs. \cite{ghd,bertini1} in  Physics {\bf 9}, 153 (2016).

\bibitem{Berkeley_bbh} V.B. Bulchandani, R. Vasseur, C. Karrasch, J.E. Moore, ``Bethe-Boltzmann Hydrodynamics and Spin Transport in the XXZ Chain'', Phys. Rev. B 97, 045407 (2018).

\bibitem{GHDdiffusion}  J. De Nardis, D. Bernard, B. Doyon, ``Hydrodynamic Diffusion in Integrable Systems'', arXiv:1807.02414.

\bibitem{takato_proof} D.-L. Vu, T. Yoshimura, ``Equations of state in generalized hydrodynamics'', arXiv:1809.03197.

\bibitem{alvise_lorenzo} A. Bastianello, L. Piroli, ``From the sinh-Gordon field theory to the one-dimensional Bose gas: exact local correlations and full counting statistics'', arXiv:1807.06869.

\bibitem{h1} A. De Luca, M. Collura, J. De Nardis, ``Non-equilibrium spin transport in integrable spin chains: persistent currents and emergence of magnetic domains", Phys. Rev. B 96, 020403 (2017).

\bibitem{h2} B. Doyon and H. Spohn, ``Dynamics of hard rods with initial domain wall state", J. Stat. Mech. 073210 (2017).

\bibitem{h3} E. Ilievski, J. De Nardis, ``Microscopic Origin of Ideal Conductivity in Integrable Quantum  Models", Phys. Rev. Lett. 119, 020602 (2017).

\bibitem{h5} L. Piroli, J. De Nardis, M. Collura, B. Bertini, M. Fagotti, ``Transport in out-of- equilibrium XXZ chains: Nonballistic behavior and correlation functions", Phys. Rev. B 96, 115124 (2017).

\bibitem{h6} E. Ilievski, J. De Nardis, ``Ballistic transport in the one-dimensional Hubbard model: the hydrodynamic approach", Phys. Rev. B 96, 081118 (2017).

\bibitem{h7} M. Fagotti, ``Higher-order generalized hydrodynamics in one dimension: The noninteracting test", Phys. Rev. B 96, 220302 (2017).

\bibitem{h8} V. B. Bulchandani, ``On classical integrability of the hydrodynamics of quantum integrable systems", J. Phys. A: Math. Theor. 50 435203 (2017).

\bibitem{h4} B. Doyon, H. Spohn, T. Yoshimura, ``A geometric viewpoint on generalized hydrodynamics", Nucl. Phys. B 926, 570 (2018).

\bibitem{h9} M. Collura, A. De Luca, J. Viti, ``Analytic solution of the domain wall non-equilibrium stationary state", Phys. Rev. B 97, 081111 (2018).

\bibitem{h10} A. Bastianello, B. Doyon, G. Watts, T. Yoshimura, ``Generalized hydrodynamics of classical integrable field theory: the sinh-Gordon model'', SciPost Phys. 4, 045 (2018).

\bibitem{footnote1} By thermal equilibrium we mean a state represented by the Gibbs ensemble at a given temperature and chemical potential.

\bibitem{chd0} S.  Stringari, ``Dynamics  of  Bose-Einstein  Condensed Gases  in  Highly  Deformed  Traps'',  Phys.  Rev.  A
58, 2385 (1998).


\bibitem{chd1} C. Menotti, S. Stringari, ``Collective oscillations of one-dimensional Bose-Einstein  gas in a time-varying trap potential and  atomic scattering length'', Phys. Rev. A 66, 043610 (2002).

\bibitem{chd2} S. Peotta, M. Di Ventra, ``Quantum shock waves and population inversion in collisions of ultracold atomic clouds", Phys. Rev. A {\bf 89}, 013621 (2014).

\bibitem{chd3} I. Bouchoule, S. Szigeti, M. Davis, K. Kheruntsyan, ``Finite-temperature Hydrodynamics for One-dimensional Bose gases: Breathing-mode Oscillations  as  a  Case  Study'', Phys. Rev. A 94, 051602(R) (2016).

\bibitem{chd4} G. De Rosi and S. Stringari, ``Hydrodynamic versus collisionless dynamics of a one-dimensional harmonically trapped Bose gas'', Phys. Rev. A {\bf 94}, 063605 (2016).

\bibitem{chd5} Y. Atas, D. Gangardt, I. Bouchoule, K. Kheruntsyan, ``Exact nonequilibrium dynamics of finite-temperature Tonks-Girardeau gases'', Phys. Rev. A {\bf 95}, 043622 (2017).

\bibitem{SMalternate} For a brief introduction of the Gross-Pitaevskii and  the classical field  models, see the Supplemental Material, which contains Refs.~\cite{Castin-coherence-2000,isabelle_twobody,jacqmin-momentum,Blakie,Cockburn-comparison-2011}.

\bibitem{Castin-coherence-2000}Y. Castin, R. Dum, E. Mandonnet, A. Minguzzi, I. Carusotto, ``Coherence properties of a continuous atom laser'', J. Mod. Opt. {\bf 47}, 2671 (2000).

\bibitem{isabelle_twobody} I. Bouchoule, M. Arzamasovs, K. V. Kheruntsyan, and D. M. Gangardt, ``Two-body momentum correlations in a weakly interacting one-dimensional Bose gas'', Phys. Rev. A {\bf 86}, 033626 (2012).


\bibitem{jacqmin-momentum} T. Jacqmin, B. Fang, T. Berrada, T. Roscilde and I. Bouchoule, ``Momentum distribution of one-dimensional Bose gases at the quasicondensation crossover: Theoretical and experimental investigation'', Phys. Rev. A, {\bf 86}, 043626 (2012).

\bibitem{Blakie} P. B. Blakie,  A. S. Bradley,  M. J. Davis,  R. J. Ballagh,   and
C. W. Gardiner, "Dynamics and statistical mechanics of ultra-cold Bose gases
using c-field techniques", Advances in Physics {\bf 57}, 363 (2008).

\bibitem{Cockburn-comparison-2011}S.P. Cockburn, A. Negretti, N. P. Proukakis, C. Henkel, ``A comparison between microscopic methods for finite temperature Bose gases'', Phys. Rev. A 83, 043619 (2011)



\bibitem{yangyang} C.N. Yang and C.P. Yang, ``Thermodynamics of a one-dimensional system of bosons with repulsive delta-function interaction'', Journal of Mathematical Physics {\bf 10}, 1115 (1969).

\bibitem{zamo} A. Zamolodchikov, ``Thermodynamic Bethe Ansatz in Relativistic Models. Scaling Three State Potts and Lee- Yang Models", Nucl. Phys. B342, 695 (1990).

\bibitem{mosselcaux} J. Mossel, J.-S. Caux, ``Generalized TBA and generalized Gibbs", J. Phys. A 45, 255001 (2012).

\bibitem{reviewcharges} E. Ilievski, M. Medenjak, T. Prosen, L. Zadnik, ``Quasilocal Charges in Integrable Lattice Systems", J. Stat. Mech. 2016, 064008 (2016).



\bibitem{these-Aisling} A. Johnson, PhD thesis (2016), https://tel.archives-ouvertes.fr/tel-01432392v1

\bibitem{phasediag} K. Kheruntsyan, D. Gangardt, P. Drummond and G. Shlyapnikov, ``Pair Correlations in a Finite-Temperature 1D Bose Gas'',  Phys. Rev. Lett. {\bf 91},040403 (2003).

\bibitem{olshanii_LL} M. Olshanii, ``Atomic Scattering in the Presence of an External Confinement and a Gas of Impenetrable Bosons'', Phys. Rev. Lett. {\bf 81}, 938 (1998).

\bibitem{Castin-Extension-2003}C. Mora and Y. Castin, "Extension of Bogoliubov theory to quasicondensates", Phys. Rev. A {\bf 67}, 053615 (2003).

\bibitem{Cazalilla-Bosonizing-2004} M. Cazalilla, "Bosonizing one-dimensional cold atomic gases", Journal of Physics B: Atomic, Molecular and Optical Physics, {\bf 37}, S1 (2004) .



\bibitem{footnote3} For each $x$, the intial distribution $\rho(x,v)$ is given  by the thermodynamic Bethe Ansatz~\cite{yangyang}.

\bibitem{SMharmonic} See the Supplemental material for a detailed discussion of why the GHD prediction is close to the CHD one for the release from a harmonic trap.

\bibitem{SMphasespace} See the Supplemental Material for plots of the distribution of quasi-particles in phase space and more details on why the double-well potential allows to discriminate between GHD and CHD.


\bibitem{SMprocedure} See the Supplemental Material for full details on our selection procedure for the GHD profiles in the case where we do not have good knowledge of the initial potential.

\bibitem{qnc} T. Kinoshita, T. Wenger, and D. S. Weiss, ``A Quantum Newton's Cradle'', Nature {\bf 440}, 900-903 (2006).

\bibitem{qnc2} Y. Tang, W. Kao, K.-Y. Li, S. Seo, K. Mallayya, M. Rigol, S. Gopalakrishnan and B. Lev, ``Thermalization near Integrability in a Dipolar Quantum Newton's Cradle'', Phys. Rev. X {\bf 8}, 021030 (2018).

\bibitem{qnc3} C. Li, T. Zhou, I. Mazets, H.-P. Stimming, Z. Zhu, Y. Zhai, W. Xiong, X. Zhou, X. Chen and J. Schmiedmayer, ``Dephasing and Relaxation of Bosons in 1D: Newton's Cradle Revisited'', arXiv:1804.01969.

\bibitem{SMperiodicity} See the Supplemental Material for plots of the distribution of quasi-particles in phase space that show that the behavior predicted by GHD is only approximately periodic.

\bibitem{footnote2} To go beyond the shock time, one would need to properly regularize the solution, for instance by considering entropy-producing diffusive terms.

\bibitem{Yang1967} C.N. Yang, ``Some Exact Results for the Many-Body Problem in one Dimension with Repulsive Delta-Function Interaction'', Phys. Rev. Lett. {\bf 19}, 1312 (1967).

\bibitem{sutherland} B. Sutherland, ``Further Results for the Many-Body Problem in One Dimension'', Phys. Rev. Lett. {\bf 20}, 98 (1968).

\bibitem{gaudin} M. Gaudin, ``Un syst\`eme a une dimension de fermions en interaction'', Physics Letters A {\bf 24}, 55 (1967).

\bibitem{batchelor} X.-W. Guan, M. Batchelor, C. Lee, ''Fermi gases in one dimension: From Bethe ansatz to experiments'', Rev. Mod. Phys. {\bf 85}, 1633 (2013).

\bibitem{whitlock} P.Wicke, S.Whitlock, N. van Druten, ``Controlling spin motion and interactions in a one-dimensional Bose gas'', arXiv:1010.4545.

\bibitem{fallani} G. Papagano et al., ''A one-dimensional liquid of fermions with tunable spin'', Nature Physics 10 (3), 198  (2014).



\end{thebibliography}
\end{document}